\documentclass[journal]{IEEEtran}
\usepackage{xcolor}
\usepackage[utf8]{inputenc}
\usepackage{amsmath,amssymb,amsfonts}
\usepackage{algorithm} 
\usepackage{epstopdf}
\usepackage{subfigure}
\usepackage{textcomp}
\usepackage{bm}
\usepackage{amsmath,amsfonts}
\usepackage{amsmath}
\usepackage{algorithmic}
\usepackage{cite}
\usepackage{array}
\usepackage[caption=false,font=normalsize,labelfont=sf,textfont=sf]{subfig}
\usepackage{xcolor}  
\usepackage{xpatch} 
\usepackage{threeparttable}
\usepackage{textcomp}
\usepackage{stfloats}
\usepackage{url}
\usepackage{verbatim}
\usepackage{graphicx}
\usepackage{booktabs}
\usepackage{threeparttable}

\def\BibTeX{{\rm B\kern-.05em{\sc i\kern-.025em b}\kern-.08em
		T\kern-.1667em\lower.7ex\hbox{E}\kern-.125emX}}
\usepackage{balance}
\begin{document}
	
	\title{\huge Heterogeneous Services in Multi-UAV Networks with No-Fly Zone and User Mobility}
	
\author{Bin Li, Senning Wan, Decheng Zhang, Yonglei Yao, and~Yu Xue

    \thanks{Bin Li, Senning Wan, Yonglei Yao and Yu Xue are with the School of Computer Science, Nanjing University of Information Science and Technology, Nanjing 210044, China (e-mail: bin.li@nuist.edu.cn; 202412492192@nuist.edu.cn).}
    
    \thanks{Decheng Zhang is with the School of Medical Information and Engineering, Bengbu Medical University, Bengbu, 233000 Anhui, China (e-mail: zdc@bbmu.edu.cn). }
}
	
	\setlength{\parskip}{0pt} 
	\maketitle
	\begin{abstract}
		In this paper, we investigate a wireless communication network supported by multiple uncrewed aerial vehicles (UAVs). To ensure reliable connectivity, UAVs are desired to approach the boundaries of designated no-fly zones (NFZs) without violating them when mobile users (MUs) enter NFZs. To support heterogeneous communication services, MUs are assigned different priority levels based on their individual communication types. Considering the practical constraints arising from user mobility, heterogeneous communication requirements, and NFZs, the objective is to maximize the weighted sum of MU communication rates, where weights correspond to MU priorities.  This is achieved by jointly optimizing trajectory planning, UAV transmission power control, and the binary association factor between UAVs and MUs. To address this challenging optimization problem under dynamic constraints, we develop a soft actor-critic algorithm with prioritized experience replay, which learns an adaptive control policy. Numerical results show that the proposed algorithm outperforms benchmark schemes in terms of convergence speed and stability. Moreover, it effectively meets diverse communication requirements, satisfies UAV energy constraints, and ensures reliable operation under NFZ constraints.
		
	\end{abstract}
	
	\begin{IEEEkeywords}
		Uncrewed aerial vehicle, no-fly zone, user mobility, heterogeneous communication, soft actor-critic.
	\end{IEEEkeywords}
	
	\section{Introduction}
	With the growing demand for high-quality wireless communications, uncrewed aerial vehicle (UAV)-based communication systems have garnered widespread attention across industry and academia. Due to the limited deployment capability of traditional terrestrial infrastructure, UAVs have emerged as a promising solution, offering high mobility and flexible deployment to dynamically adjust their positions for optimal communication coverage. With these advantages, UAVs play a vital role in applications such as autonomous driving, target localization, disaster relief, and public emergency events~\cite{art1,art2,art3}. Consequently, UAVs are envisioned to overcome the limitations of conventional wireless networks by acting as flexible communication platforms, including mobile relays and airborne base stations (BSs). In particular, UAVs can act as airborne BSs to rapidly provide wireless services in emergency or unexpected scenarios~\cite{art4,art5,10186380}.
	
   In practical deployments, UAV-assisted communication systems often operate in environments with no-fly zones (NFZs).  NFZs are typically designated for public safety, privacy, or national security reasons. In particular, mobile users (MUs) may enter designated NFZs.  For instance, in disaster recovery, UAVs must provide connectivity to affected MUs while avoiding hazardous areas such as fire zones or chemical leakages. Similarly, in urban surveillance, UAVs are frequently constrained near airports or government facilities.  In such cases, MUs are often involved in different communication types, which leads to heterogeneous service demands. Therefore, it is essential to develop flexible communication frameworks that can adapt to heterogeneous requirements~\cite{art8,art9}. In light of these challenges, this paper jointly considers user mobility and communication heterogeneity under NFZ constraints in both system modeling and algorithm design.

	\subsection{Previous Work}
	\textit{1) UAV Communications:} To fully exploit the potential of UAV-assisted communication, numerous studies have focused on UAV trajectory optimization under various communication scenarios. Proper trajectory planning can effectively reduce the communication distances between UAVs and MUs, thereby significantly increasing the transmission rate. Given the complexity of the problem and the need for robustness, convex optimization techniques have become a powerful tool for UAV trajectory planning~\cite{cvx1,cvx2,cvx3}. In~\cite{cvx1}, Yu \textit{et al.} proposed employing a reconfigurable intelligent surface (RIS) to mitigate interference by alternately optimizing the UAV trajectory and RIS passive beamforming. Similarly, Jin \textit{et al.}~\cite{cvx2} investigated the throughput maximization problem by applying successive convex optimization techniques. Furthermore, Katwe \textit{et al.}~\cite{cvx3} investigated the throughput maximization problem in a multi-UAV communication system by applying successive convex approximation and a brute-force search algorithm. Recently, deep reinforcement learning (DRL) has received considerable attention, particularly for UAV trajectory planning under uncertain and dynamic communication conditions~\cite{rl1,SAC,rl2}. Liu \textit{et al.}~\cite{rl1} proposed a multi-UAV control algorithm based on the deep deterministic policy gradient (DDPG) framework to improve energy efficiency. Zhao \textit{et al.}~\cite{SAC} adopted a distributionally robust reinforcement learning (RL) approach to jointly optimize UAV trajectories as well as active and passive beamforming.  Cao \textit{et al.}~\cite{rl2} proposed a centralized training multi-agent soft actor–critic with the branch-and-cut algorithm to maximize system throughput and coverage, where each UAV and user is treated as an individual learning agent.
	
	 \textit{2) NFZ Constraints:} In practical applications, UAV flight paths are often constrained by the geographic environment and therefore must be carefully designed. Several studies have incorporated NFZ constraints into UAV trajectory design~\cite{nfz1,nfz2,nfz3,nfz4}. In particular, Wang \textit{et al.}\cite{nfz1} considered constraints such as limited power resources, user quality-of-service (QoS) requirements, and NFZ avoidance, with the objective of maximizing the system sum-rate. However, their work was limited to a single-UAV scenario. In\cite{nfz2}, Xu \textit{et al.} investigated a practical multi-user multiple-input single-output UAV communication system under NFZ constraints. In~\cite{nfz3}, Li \textit{et al.} studied resource allocation and trajectory design for secure UAV communication systems in the presence of cylindrical NFZs, with the objective of maximizing the average minimum secrecy rate per user. Heo \textit{et al.}~\cite{nfz4} proposed a joint optimization framework for UAV trajectory and communication resource allocation that completely avoids NFZs over the entire continuous trajectory. However, these studies neglect how UAV trajectories should be adaptively adjusted to maintain communication continuity when MUs move into NFZs.
	
	\textit{3) Heterogeneous Requirements:} Considering that MUs have diverse service demands in practical scenarios, several recent studies have addressed heterogeneous user requirements~\cite{HC1,HC2,HC3}. Zhang \textit{et al.}\cite{HC1} aimed to maximize the overall sum-rate of all users under constraints on each user’s QoS requirements and the total available resources. Wang \textit{et al.}\cite{HC2} categorized users into QoS-sensitive and QoS-tolerant groups according to their data rate requirements. In~\cite{HC3}, Yu \textit{et al.} investigated the joint optimization of three-dimensional (3D) UAV deployment and UAV selection for machine-type devices with heterogeneous data requirements, and proposed an air–ground collaborative deployment strategy to optimize user satisfaction. The distinctions between our proposed scheme and the relevant existing schemes are summarized in Table~\ref{Scheme_Comparison}. For brevity, the term UAV is used throughout this paper to refer to UAV BSs. 
	 
\begin{table*}[t]
\centering
\begin{threeparttable}
\caption{Differences in Our Scheme Compared to the Most Relevant Schemes}
\label{Scheme_Comparison}
\begin{tabular}{|c|c|c|c|c|c|c|}
\hline
Schemes & Multi-UAV & NFZ Constraint & User Mobility & Heterogeneous Service & Objective Function & Learning Methodology  \\
\hline
{\cite{cvx2}}  & -- & -- & -- & -- & Sum-Rate & CVX\tnote{1}  \\
{\cite{cvx3}}  & \checkmark & -- & -- & -- &  Sum-Rate & CVX \\
{\cite{rl1}}  & \checkmark & -- & -- & -- &  Energy Efficiency & DRL  \\
{\cite{nfz1}}  & -- & \checkmark & \checkmark & -- & Sum-Rate &  CVX \\
{\cite{nfz4}} & -- & \checkmark & -- & -- & Minimum Average Rate & CVX  \\
{\cite{HC3}} & \checkmark & -- & -- & \checkmark & User Satisfaction & Hierarchical Game  \\
\hline
Our Scheme & \checkmark & \checkmark & \checkmark & \checkmark & Priority Rate  & DRL \\
\hline
\end{tabular}

\begin{tablenotes}
\footnotesize
\item[1] CVX is a MATLAB-based package for specifying and solving convex programs.
\end{tablenotes}

\end{threeparttable}
\end{table*}
	
	\subsection{Motivations and Contributions}
	Although most existing studies have considered the restrictions imposed by NFZs on UAV flight, they largely overlook how UAVs should dynamically adapt their communication strategies when MUs enter these restricted areas. This reveals a gap between NFZ-aware trajectory design and communication adaptation to user mobility and heterogeneity. To bridge this gap, the main contributions of this work are summarized as follows:
	\begin{itemize}
    \item We consider a multi-UAV-enabled wireless communication scenario, where UAVs act as airborne BSs. Considering user mobility and heterogeneous service requirements, MUs are classified into different priority levels according to their communication types. In addition, cylindrical NFZs are introduced, and their impact on UAV trajectory planning is analyzed. For system performance optimization, we formulate a joint problem to maximize the sum of priority rates across all MUs by jointly optimizing transmission power allocation, UAV trajectory planning, and binary UAV-MU association variables.  

    \item We formulate the problem of maximizing the sum of priority rates as a Markov decision process (MDP). To address the challenges posed by channel dynamics, heterogeneous service requirements, and user mobility, we employ a DRL algorithm based on the actor–critic framework, namely the soft actor–critic with prioritized experience replay (SAC-PER), to facilitate optimal decision-making.

	\item Through simulations under various scenarios, we verify that, under the proposed constraints, UAVs can adaptively approach MUs located within NFZs to provide reliable communication while strictly avoiding NFZ boundaries. Consequently, the proposed scheme achieves higher efficiency and priority rates than benchmark schemes in all considered environments.

	\end{itemize}
	
	\subsection{Organization}
	The rest of this paper is organized as follows. Section II presents the system model. Section III formulates the joint trajectory, power, and association optimization problem. Section IV develops the MDP-based SAC-PER algorithm. Section V presents the simulation results, and Section VI concludes the paper.

	\section{System Model}\label{s:sys}
	This section introduces the multi-UAV assisted communication system. The network model, UAV mobility model, communication model, and user mobility model are described in detail. Table~\ref{tab:Symbol} presents all the important symbols.
        	    \begin{table}[htbp]
    \renewcommand{\arraystretch}{1.3}
    \centering
    \caption{IMPORTANT NOTATIONS OF OUR ARTICLE}
    \label{tab:Symbol}
    \begin{tabular}{|>{\centering\arraybackslash}p{0.1\textwidth}|>{\centering\arraybackslash}p{0.33\textwidth}|}
        \hline
        \textbf{Symbol} & \textbf{Meaning} \\
        \hline
        $N$ & The number of UAVs \\
        \hline
        $K$  & The number of MUs\\
        \hline
        $I$ & The number of NFZs \\
        \hline
        $Z$ & The number of communication types \\
        \hline
        $T$ & The number of time slots \\
        \hline
        $\mathbf{q}_n[t]$ &  The coordinate of UAV $n$ at time slot $t$\\
        \hline
        $\mathbf{u}_k[t]$ &  The coordinate of MU $k$ at time slot $t$\\
        \hline
        $\mathbf{q}_{\text{NFZ}}^i$,${R}_{\text{NFZ}}^i$ &  The center and radius of the NFZ $i$ \\
        \hline       
         $d_{\text{min}}$&  Minimum safe distance between UAVs \\
        \hline
        ${d}_{\text{NFZ}}$ &  NFZ boundary safety distance  \\
        \hline
        $E_n^\text{fly}[t]$ & Flight energy consumption of UAV $n$\\
        \hline
        $E_n^{\text{res}}[t]$ & Residual energy of UAV $n$ \\
        \hline
        $E_n^{\text{UAV}}$ & Battery capacity of UAV $n$ \\
        \hline
        $p_{n,k}[t]$ & The transmission power of UAV $n$ to MU $k$\\
        \hline
        $\xi_{k,z}[t]$ & The matching indicator between MU $k$ and communication type $z$ \\
        \hline
        $\zeta_z$ & Priority level of communication type $z$\\
        \hline
        $\alpha_{n,k}[t]$ &  The association factor between UAV $n$ and MU $k$\\
        \hline
        $R_{n,k}[t]$ & The transmission rate from UAV $n$ to MU $k$ \\
        \hline
    \end{tabular}
\end{table}

	\subsection{Network Model}
\begin{figure}[t]
		\centering
		\includegraphics[width=\columnwidth]{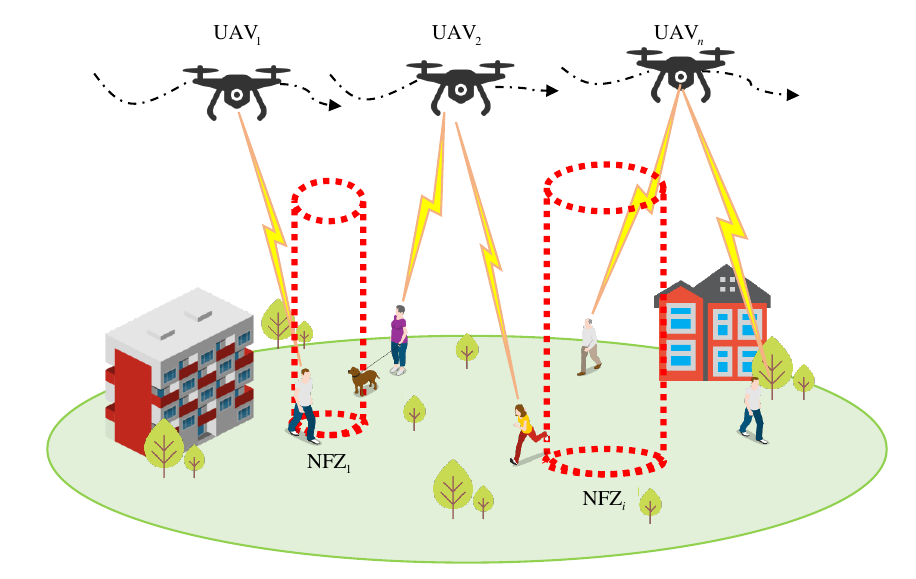}
		\caption{Illustration of a multi-UAV-assisted communication system.}
		\label{fig:sys-model}
	\end{figure}

	In this part, we investigate a UAV-assisted communication scenario that considers $I$ circular cylindrical NFZs. As illustrated in Fig.~\ref{fig:sys-model}, UAVs are explicitly prohibited from entering these NFZs to ensure safe and regulatory-compliant operation. The proposed system consists of $N$ UAVs and $K$ MUs. Each UAV is equipped with an aerial BS to provide communication services to MUs, which are spatially distributed across different locations. Let $\forall n \in \mathcal{N} \triangleq \{1,2,\dots,N\}$ and  $\forall k \in \mathcal{K} \triangleq \{1,2,\dots,K\}$ denote the sets of UAV and MU indices, respectively. The UAV operates within the feasible region $\mathcal{Q} = \{ (x,y) \in \mathbb{R}^2 \;\mid\; 0 \leq x \leq x_{\max}, \; 0 \leq y \leq y_{\max} \}$, where \(x_{\max}\) and \(y_{\max}\) represent the maximum boundaries of the flying area. The center and radius of the $i$-th NFZ are denoted by $\mathbf{q}_{\text{NFZ}}^i = \left( x_{\text{NFZ}}^i, y_{\text{NFZ}}^i \right)^T$ and $R_{\text{NFZ}}$, respectively, where $\forall i \in \mathcal{I} \triangleq \{1,2,\dots,I\}$. A complete service cycle of each UAV is denoted by $L$, which is uniformly divided into $T$ equal and non-overlapping time slots, each of duration $\delta$, such that $\delta = \frac{L}{T}$. Let $\forall t \in \mathcal{T} \triangleq \{1,2,\dots,T\}$ denote the set of time slots.

	\subsection{UAV Mobility Model}

    \begin{figure}[t]
		\centering
		\subfigure[Scenario 1: MU is outside an NFZ.]{
			\includegraphics[width=0.22\textwidth]{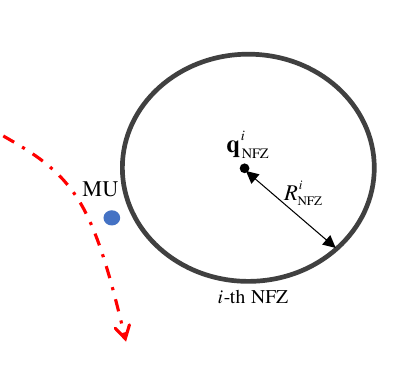}
		}
		\subfigure[Scenario 2: MU inside an NFZ.]{
			\includegraphics[width=0.22\textwidth]{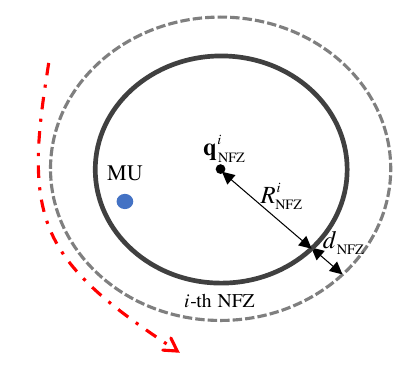}
		}
		\caption{UAV trajectories in different scenarios.}
		\label{fig:uav_trajectories}
	\end{figure}
    
	 	To ensure generality, we establish a 3D Cartesian coordinate system. The position of $k$-th MU at time slot $t$ is given by $\mathbf{u}_k[t] = \left( x^u_k[t], y^u_k[t]\right)^T$. Similarly, the position of the $n$-th UAV is denoted by $\mathbf{q}_n[t] = \left( x_n[t], y_n[t] \right)^T$. To conserve energy and avoid unnecessary consumption due to frequent altitude adjustments, all UAVs are considered to maintain an altitude fixed at $H$ throughout the operation. Each UAV initially takes off from a randomly selected position. In addition, we consider $I$ non-overlapping NFZs, which are randomly distributed and modeled as vertical cylindrical regions with a fixed altitude $H_{\text{NFZ}}$, satisfying $H \leq H_{\text{NFZ}}$. Considering real-world scenarios, the speed and acceleration of the UAVs should satisfy constraints 
	\begin{equation}
		\|\mathbf{v}_n[t]\| \leq v_{\text{max}},  \forall n \in \mathcal{N}, t \in \mathcal{T},
		\label{eq:velocity_constraint}
	\end{equation}
	\begin{equation}
		\|\mathbf{a}_n[t]\| \leq a_{\text{max}},  \forall n \in \mathcal{N},t \in \mathcal{T},
		\label{eq:acc_constraint}
	\end{equation}
	where $v_{\text{max}}$ denotes the maximum speed, and $a_{\text{max}}$ represents the maximum acceleration. The trajectory of the UAVs should satisfy
\begin{equation}
v_{n}[t+1] = v_{n}[t] + a_{n}[t]\delta,
\label{eq:velocity_update}
\end{equation}
	\begin{equation}
		\mathbf{q}_n[t+1] = \mathbf{q}_n[t] + \mathbf{v}_n[t] \delta + \frac{1}{2} \mathbf{a}_n[t] \delta^2,
        \label{eq:Trajectory_control}
	\end{equation}
	\begin{equation}
		\|\mathbf{q}_n[t] - \mathbf{q}_{n'}[t]\|^2 \geq d_{\text{min}}^2, \forall n,n' \in \mathcal{N},\; n \ne n', 
		\label{eq:security_constraint}
	\end{equation}
	where $d_{\text{min}}$ is the minimum safety distance for UAVs flight. In practical scenarios, due to safety concerns, UAVs are not allowed to fly arbitrarily. Specifically, there are boundary  and NFZ constraints. Therefore, considering these limitations, the UAV trajectory constraints can be expressed as
	\begin{equation}
		\mathbf{q}_n[t] \in \mathcal{Q}, \forall n \in \mathcal{N},
		\label{eq:x_constraint}
	\end{equation}
	\begin{equation}
		\|\mathbf{q}_n[t] - \mathbf{q}_{\text{NFZ}}^i\|^2 \geq (R_{\text{NFZ}})^2,  \forall n \in \mathcal{N},i \in \mathcal{I}.
		\label{eq:NFZ_constraint}
	\end{equation}

    Specifically, when an MU is located outside an NFZ, the UAV only needs to ensure that (\ref{eq:NFZ_constraint}) is satisfied. However, if the MU moves into the NFZ at a certain time slot $\|\mathbf{u}_k[t]-\mathbf{q}_{\text{NFZ}}^i\|^2 < (R_{\text{NFZ}})^2$ , the UAV trajectory should be adjusted to maintain the communication quality. To address this issue, we introduce a safety distance $d_{\text{NFZ}}$ from the NFZ boundary. In this case, the UAV $n$ is required to satisfy

\begin{equation}
\begin{split}
    (R_\text{NFZ}^{i})^{2} &< ||\mathbf{q}_{n}[t]-\mathbf{q}_\text{NFZ}^{i}||^{2} \le (R_\text{NFZ}^{i}+d_\text{NFZ})^{2}, \\
    &\text{if } \alpha_{n,k}[t]=1 \text{ and } ||\mathbf{u}_{k}[t]-\mathbf{q}_\text{NFZ}^{i}||^{2} < (R_\text{NFZ}^{i})^{2}.
\end{split}
\label{eq:constraint_7}
\end{equation}

	Fig.~\ref{fig:uav_trajectories} illustrates the trajectory evolution of the UAV under different scenarios. Furthermore, to prevent a UAV from crossing an NFZ between discrete time steps, each time slot $\delta$ is practically divided into $M$ micro-slots. Spatial constraints, such as \eqref{eq:constraint_7}, are evaluated at these micro-intervals to strictly guarantee continuous safety.
    
    The total energy consumption of a UAV consists of two main components. The first is the communication-related energy, which arises from radiation, signal processing, and other circuitry operations. The second is the propulsion energy, which is required to keep the UAV aloft and to support its mobility when necessary. It is worth noting that, in practice, the communication-related energy is usually negligible compared to the propulsion energy of UAVs~\cite{7888557} , and is therefore ignored in this paper. When the UAVs are flying, the propulsion power consumption $P_n^{\text{fly}}[t]$ is modeled as
	\begin{align}
		P_n^{\text{fly}}[t] = & \frac{1}{2} d_0 \rho_1 g_0 A_0 \|\mathbf{v}_n[t]\|^3 
		+ P_1 \left(1 + \frac{3 \|\mathbf{v}_n[t]\|^2}{U_{\text{tip}}^2} \right) \notag \\
		& + P_2 \left(1 + \frac{\|\mathbf{v}_n[t]\|^4}{4v_0^4} - \frac{\|\mathbf{v}_n[t]\|^2}{2v_0^2} \right)^{\frac{1}{2}},
	\end{align}
	where $P_1$ and $P_2$ denote the power consumed by the UAV's rotor blades and the induced power during hovering, respectively. Additionally, $v_0$ represents the average rotor velocity, $\rho_1$ refers to the air density, $U_{\text{tip}}$ indicates the tip speed of the rotor blades, $d_0$ refers to the fuselage drag ratio, $A_0$ is the rotor area, and $g_0$ signifies the solidity of the rotor blade. Therefore, the flight energy consumption of the UAV $n$ at time slot $t$ can be calculated as
	\begin{equation}
		E_n^{\text{fly}}[t] = P_n^{\text{fly}}[t] \cdot \delta.
	\end{equation}

	\subsection{Communication Model}
	In this section, we assign priorities to MUs based on their communication types to accommodate diverse communication requirements. The communication types are indexed by $\forall z \in \mathcal{Z} \triangleq \{1, 2, \dots, Z\}$, with the corresponding priority level denoted by $\zeta_z$. The matching indicator between MU $k$ and communication type $z$ at time slot $t$ is represented by
	\begin{equation}
		\xi_{k,z}[t] \in \{0, 1\}, \forall k \in \mathcal{K}, z \in \mathcal{Z},
		\label{eq:xi1}
	\end{equation}
	\begin{equation}
		\sum_{z=1}^{Z} \xi_{k,z}[t] = 1,  \forall k \in \mathcal{K},
		\label{eq:xi2}
	\end{equation}
	when $\xi_{k,z}[t] = 1$, it indicates that MU $k$ belongs to communication type $z$; otherwise, $\xi_{k,z}[t] = 0$. Constraint~\eqref{eq:xi2} represents that each MU can belong to only one communication type. To optimize communication performance, the system must reasonably assign UAVs to MUs for communication. The binary association factor between UAV $n$ and MU $k$ at time slot $t$ is defined as
	\begin{equation}
		\alpha_{n,k}[t] \in \{0, 1\},  \forall n \in \mathcal{N}, k \in \mathcal{K}, t \in \mathcal{T},
		\label{eq:alpha1}
	\end{equation}
	\begin{equation}
		\sum_{n=1}^{N} \alpha_{n,k}[t] = 1,  \forall k \in \mathcal{K}, t \in \mathcal{T},
		\label{eq:alpha2}
	\end{equation}
	when $\alpha_{n,k}[t] = 1$, it denotes that MU $k$ is connected to UAV $n$ at time slot $t$, allowing communication to take place; otherwise, $\alpha_{n,k}[t] = 0$. Constraint~\eqref{eq:alpha2} ensures that each MU is associated with exactly one UAV. In practical scenarios, the communication channels between MUs and UAVs are typically affected by various obstacles, including trees and buildings. Consequently, the channel link between UAV $n$ and MU $k$ is modeled as a probabilistic model that incorporates both line-of-sight (LoS) and non-line-of-sight (NLoS) components. Accordingly, the estimated channel state information (CSI) between UAV $n$ and MU $k$ at time slot $t$ is given by
	\begin{equation}
		h_{n,k}[t] = \sqrt{\rho d_{n,k}^{-\beta}[t]} 
		\left( \sqrt{\frac{\varsigma}{1+\varsigma}} \, \bar{h}_{n,k}^{\text{LoS}}[t] 
		+ \sqrt{\frac{1}{1+\varsigma}} \, \tilde{h}_{n,k}^{\text{NLoS}}[t] \right),
	\end{equation}
	where $\rho$ represents the channel power gain at the reference distance,  $\beta$ is the path loss exponent, $d_{n,k}[t] = \sqrt{\left\| \mathbf{q}_n[t] - \mathbf{u}_k[t] \right\|^2+ H^2}$ represents the distance between UAV $n$ and MU $k$ at time slot $t$, and $\varsigma$ denotes the Rician factor. The NLoS channel coefficient between UAV \( n \) and MU \( k \), denoted as \( \tilde{h}_{n,k}^{\text{NLoS}}[t] \), is modeled as a random vector with independent and identically distributed complex Gaussian variables with zero mean and unit variance, i.e., \( \tilde{h}_{n,k}^{\text{NLoS}}[t] \sim \mathcal{CN}(0, 1) \), and \( \bar{h}_{n,k}^{\text{LoS}}[t] \) represent the LoS components. We assume that orthogonal frequency-division multiple access is employed, which inherently eliminates intra-UAV interference. Thus, only inter-UAV interference is considered. Accordingly, the resulting signal-to-interference-plus-noise ratio (SINR) can be expressed as
\begin{equation}
    \Upsilon_{n,k}[t]=\frac{|h_{n,k}[t]|^{2}p_{n,k}[t]}{\sum_{j\ne n}^{N}|h_{j,k}[t]|^{2}\sum_{l=1}^{K}\alpha_{j,l}[t]p_{j,l}[t]+\sigma^{2}},
    \label{eq:SINR}
\end{equation}
	where $\sigma^2$ denotes the noise power, and $p_{n,k}[t]$ represents the transmission power of UAV $n$ to MU $k$ at time slot $t$. The term $\sum_{\ell=1}^{K}\alpha_{j,\ell}[t]p_{j,\ell}[t]$ denotes the total transmit power of interfering UAV $j$ allocated to its associated MUs. Therefore, the transmission rate from UAV $n$ to MU $k$ at time slot $t$ is given by
	\begin{equation}
		R_{n,k}[t] = B_0 \log_2 \left(1 + \Upsilon_{n,k}[t] \right),
		\label{eq:prior-rate}
	\end{equation}
	where $B_0$ represents the channel bandwidth. Therefore, the sum of priority rates across all MUs is defined as
	\begin{equation}
	\bar{R}[t] = \sum_{n=1}^{N} \sum_{k=1}^{K} \sum_{z=1}^{Z}  \xi_{k,z}[t] \zeta_z \alpha_{n,k}[t] R_{n,k}[t].
\end{equation}

   \subsection{User Mobility Model}
	To better adapt to realistic scenarios, this paper adopts the recurrent self-similar Gauss-Markov (RSSGM) model to characterize user movement~\cite{GMM}. The RSSGM model effectively and accurately fits movement trajectories by combining historical states with a Gaussian distribution, adapting to different environments while providing efficient computation and prediction capabilities. In addition, Gaussian noise is introduced to smooth the motion trajectories, making them more similar to typical mobility patterns observed in everyday human activities.
	
	\textit{1) Speed Update:} The speed of MU $k$ is updated intermittently over time to reflect realistic movement behavior. The update rule is given by
	\begin{equation}
		v_k[t] = \eta v_k[t-1] + (1 - \eta) v_{\text{rand}} + \sqrt{1 - \eta^2} \, \varpi_{\text{noise}},
	\end{equation}
	where $v_k[t]$ denotes the speed of MU $k$ at time slot $t$. The parameter $\eta$, referred to as the speed relevance factor, satisfies $0 \leq \eta \leq 1$, and governs the degree of speed persistence. A larger $\eta$ indicates stronger inertia, resulting in smoother and more continuous movement; conversely, when $\eta = 0$, the velocity becomes memoryless, resembling a pure random walk. The term $v_{\text{rand}}$ represents a randomly generated speed uniformly sampled from the interval $\left[v_u^{\text{min}}, v_u^{\text{max}}\right]$, while $\varpi_{\text{noise}}$ denotes zero-mean Gaussian noise with variance $\sigma^2$, i.e., $\varpi_{\text{noise}} \sim \mathcal{N}(0, \sigma^2)$. The factor $\sqrt{1 - \eta^2}$ normalizes the impact of noise based on the chosen persistence level.
	
	\textit{2) Direction Update:} The moving direction of MU $k$ also evolves over time and is modeled using a Markov process to reflect temporal continuity. The direction is updated as
	\begin{equation}
		\theta_k[t] = \theta_k[t-1] + \text{clip}(\mathcal{N}(0, \sigma_\theta^2), -\Delta \theta_{\text{max}}, \Delta \theta_{\text{max}}),
		\label{eq:user_dir}
	\end{equation}
	where $\theta_k[t]$ denotes the movement direction of MU $k$ at time slot $t$. The term $\mathcal{N}(0, \sigma_\theta^2)$ represents a Gaussian random variable with zero mean and variance $\sigma_\theta^2$, used to introduce randomness in directional changes. To prevent abrupt or physically unrealistic turns, the function $\text{clip}(\cdot)$ restricts the perturbation to the interval $[-\Delta \theta_{\mathrm{max}}, \Delta \theta_{\mathrm{max}}]$, where $\Delta \theta_{\mathrm{max}} = \arctan \left( v_u^{\mathrm{max}} \right)$. This constraint ensures that direction variations remain within a reasonable range, producing smoother trajectories that align with human mobility patterns.

	\textit{3) User Position Update:} Based on the updates of speed and direction, the position is expressed as
	\begin{equation}
		\mathbf{u}_k[t+1] = \mathbf{u}_k[t] + v_k[t] \cdot \mathbf{d}_k[t] \cdot \delta,
		\label{eq:position_update}
	\end{equation}
	where $\mathbf{d}_k[t] = (\sin(\theta_k[t]), \cos(\theta_k[t]))^T$ represents the direction vector of the $k$-th MU at time slot $t$. This RSSGM-based mobility framework captures both spatial continuity and temporal correlation, which makes it well-suited for simulating realistic movement patterns in dynamic wireless communication scenarios.

\subsection{Limitations and Extensions}
For analytical tractability, the system model adopts several simplifying assumptions, including a fixed UAV altitude and perfect CSI. In practice, altitude control and imperfect CSI can be accommodated by augmenting the state with altitude dynamics and CSI uncertainty, and by employing robust or risk-sensitive control strategies.
	
	\section{Problem Formulation}
    
We aim to optimize the UAV trajectories $\mathbf{q}_n[t]$ over the given time horizon. As indicated by \eqref{eq:Trajectory_control}, the UAV trajectories are governed by the acceleration dynamics. Therefore, we introduce the acceleration vector $\mathbf{a}_n[t]$ as the decision variable. Based on equation~\eqref{eq:prior-rate}, we further consider a joint optimization framework that involves the UAV acceleration $\mathbf{A} \triangleq \{\mathbf{a}_{n}[t], \forall n \in \mathcal{N}, t \in \mathcal{T}\}$, the UAV power allocation $\mathbf{P} \triangleq \{p_{n,k}[t], \forall n \in \mathcal{N}, k \in \mathcal{K}, t \in \mathcal{T}\}$, and the binary association variables between MUs and UAVs $\bm{\alpha} \triangleq \{\alpha_{n,k}[t], \forall n \in \mathcal{N}, k \in \mathcal{K}, t \in \mathcal{T}\}$, with the objective of maximizing the sum of priority rates across all MUs. Accordingly, the optimization problem can be formulated as follows:

	\begin{subequations} 
		\begin{align}     
			{ \max_{\mathbf{A}, \mathbf{P}, \bm{\alpha}} }& \quad {\sum_{t=1}^{T} \bar{R}[t]} \label{eq:a} \\     
			\text{s.t.} & \quad 
			\eqref{eq:velocity_constraint}, \eqref{eq:acc_constraint}, \eqref{eq:security_constraint}, \eqref{eq:x_constraint}, \eqref{eq:NFZ_constraint}, 
            \eqref{eq:constraint_7}, \eqref{eq:alpha1}, \eqref{eq:alpha2}, \label{eq:b} \\    
			& \quad 0 \leq p_{n,k}[t] \leq \alpha_{n,k}[t] P_n^{\max}, \forall n \in \mathcal{N}, k \in \mathcal{K} \label{eq:c}, \\     
			& \quad \sum_{k=1}^{K}\alpha_{n,k}[t] p_{n,k}[t] \leq P_n^{\max}, \forall n \in \mathcal{N}, t \in \mathcal{T} \label{eq:d}, \\     
			& \quad  E_n^{\text{res}}[t] \geq 0, \forall n \in \mathcal{N}, t \in \mathcal{T} \label{eq:e}. 
		\end{align} 
		\label{eq:sum} 
	\end{subequations} 
    
	Constraints \eqref{eq:velocity_constraint} and \eqref{eq:acc_constraint} restrict the speed and acceleration of the UAV movement. Constraint \eqref{eq:security_constraint} ensures a minimum safety distance between UAVs to prevent collisions. Constraint \eqref{eq:x_constraint} defines the rectangular boundaries of the UAV flight area. Constraint \eqref{eq:NFZ_constraint} represents the limitation imposed by NFZs on UAV flight paths. Constraint \eqref{eq:constraint_7} imposes a specific flight trajectory restriction on the associated UAV when its served MU moves into an NFZ.
    Constraints \eqref{eq:alpha1} and \eqref{eq:alpha2} ensure that each MU can be associated with at exactly one UAV. Constraint \eqref{eq:c} couples the transmit power with the UAV-MU association variable, ensuring that $p_{n,k}[t]=0$ when MU $k$ is not associated with UAV $n$, while allowing power allocation only for associated UAV-MU pairs. Constraint \eqref{eq:d} limits the transmission power of the UAV $n$ in each time slot, which is limited by the maximum allowed power $P_n^{\max}$.  Finally, constraint \eqref{eq:e} requires the remaining energy of each UAV to be nonnegative, where \( E_n^{\text{res}}[t] = E_n^{\text{UAV}} - \sum_{s=1}^{t} E_n^{\text{fly}}[s] \).

	\section{Problem Solution}\label{pro:s}
Problem~\eqref{eq:sum} formulates a mixed-integer nonlinear programming problem, which is inherently non-deterministic polynomial (NP)-hard and thus exhibits significant computational complexity. Specifically, this problem involves two categories of decision variables: the continuous vectors, namely the power vector $\mathbf{P}$ and the acceleration vector $\mathbf{A}$, and the discrete binary indicator variable $\boldsymbol{\alpha}$. Moreover, the nonlinear characteristics of the objective function together with constraints \eqref{eq:d} and~\eqref{eq:e} further exacerbate the problem complexity. Although various convex optimization techniques can be applied to transform the problem into a convex form for tractable solutions, the resulting computational overhead severely degrades the system performance, which is unacceptable in UAV-assisted communication scenarios.

SAC is a DRL algorithm built upon the policy gradient framework. Unlike traditional optimization methods such as linear programming, DRL does not rely on complex mathematical modeling and is therefore particularly suitable for addressing NP-hard problems such as~\eqref{eq:sum}. Motivated by this, we adopt SAC-PER to solve the proposed problem, as it offers high sample efficiency and strong exploration capability in high-dimensional continuous control tasks. These properties make SAC particularly effective for optimizing continuous decision variables and allow the incorporation of discrete components through appropriate action-space design. In addition, the replay buffer mechanism in SAC enables efficient reuse of historical experiences, thereby improving adaptability to dynamic channel variations and time-varying system states.  
	
	\subsection{Modeling of MDP}
	Before delving into the design of RL algorithms, we first model the formulated problem using the MDP framework. The MDP is formulated as a quintuple $(S, A, P, R, \gamma)$, where $S$ and $A$ denote the state and action spaces, respectively,  $P$ denotes the state transition probability, $R$ specifies the reward function, and $\gamma \in [0,1]$ is the discount factor. Let $\rho_{\pi}$ denote the state-action trajectory distribution following the policy $\pi$. At each time slot, the agent selects an action $a_t \in A$ following the policy $\pi$, which transforms the state $s_t \in S$ and obtains the reward $r_t = r(s_t, a_t)$ from the environment. To maximize the sum of priority rates, the design of the state, action, and reward functions is of critical importance, and their detailed descriptions are provided below.
	
	\textit{1) State:}
	The environmental state $s_t$ observed by the agent consists of the following parameters:
	\begin{itemize}
		\item[a)] The position of UAVs \( \mathbf{q}_n[t] \).
		\item[b)] The position of MUs \( \mathbf{u}_k[t] \).
		\item[c)] The residual energy of UAVs at time slot \(t\) is denoted \(E_n^{\text{res}}[t]\).
		\item[d)] The speed of UAVs, denoted as  \( \mathbf{v}_n[t] \).
		\item[e)] The communication type matching factor $\xi_{k,z}[t]$.
	\end{itemize}
	
	\textit{2) Action:}
	In reinforcement learning, the agent interacts with the environment by outputting actions at each time slot. Therefore, the optimization process is essentially equivalent to learning a policy that maps the observed states to actions, which makes the optimization variables identical to the action space. Based on the received state $s_t$, the agent selects the corresponding action $a_t$, with the action parameters specified as follows:
	\begin{itemize}
		\item[a)] The acceleration magnitude and direction of UAVs, denoted as \( \mathbf{a}_n[t] = \left[ \|\mathbf{a}_n[t]\|, \phi_n[t] \right] \), where \( \|\mathbf{a}_n[t]\| \) represents the acceleration and \( \phi_n[t] \) represents the change in the direction of UAVs.
		\item[b)] The transmission power from UAVs to MUs, denoted as \( p_{n,k}[t] \).   
		\item[c)] The communication association indicator for MUs is denoted as \( \alpha_{n,k}[t] \). According to the policy model, the binary UAV-MU association variable for each MU is determined as
        \begin{equation}
        \alpha_{n,k}[t] =
        \begin{cases}
        1, & n = \arg \max\limits_{j} \hat{\alpha}_{j,k}[t], \\
        0, & \text{otherwise},
         \end{cases}
      \end{equation}
      where $\hat{\alpha}_{n,k}[t]$ denotes the continuous output score of the policy model. This guarantees that each MU is associated with only one UAV at each time slot.
	\end{itemize} 
	
	\textit{3) Reward:}
    According to \eqref{eq:sum}, the objective of designing the UAV trajectory, transmit power, and association decisions is to maximize the sum of priority rates over the time slot $t$. Since the reward in the guided learning process should be consistent with this objective, we simply include the instantaneous sum-rate of all MUs at each time slot, given by $\bar{R}[t]$ in the reward function. The reward function is defined to account for various penalty scenarios, the reward $r_t$ is calculated as
	\begin{equation}
		r_t = 
		\begin{cases} 
			\bar{R}[t] - P_1, & \text{if} ~\mathbf{q}_n[t] \notin \mathcal{Q}, \\
			\bar{R}[t] - P_2, & \text{if } \|\mathbf{q}_n[t] - \mathbf{q}_{\text{NFZ}}^i\|^2 < (R_{\text{NFZ}})^2, \\
			\bar{R}[t] - P_3, & \text{if } \|\mathbf{q}_n[t] - \mathbf{q}_{n'}[t]\|^2 < d_{\min}^2, \\
			\bar{R}[t] + P_4, & \text{if constraint~(\ref{eq:constraint_7}) is satisfied, }\\
            \bar{R}[t] + w E_n^{\text{res}}[t], & \text{if } t<T ~\text{and}~E_n^{\text{res}}[t] < 0, 
		\end{cases}
	\end{equation}
where boundary penalty is expressed as
\begin{equation}
P_{1} = \kappa_{1} \sum_{n=1}^{N} d^{2}\!\big(\mathbf{q}_{n}[t], \mathcal{Q}\big),
\end{equation}
where $d(\mathbf{q}_n[t], \mathcal{Q})$ denotes the minimum Euclidean distance from the UAV position $\mathbf{q}_n[t]$ to the region $\mathcal{Q}$. The NFZ penalty is defined as
\begin{equation}
P_{2} = \kappa_{2} \sum_{i=1}^{I} \sum_{n=1}^{N} D_{ni}^{2},
\end{equation}
where $D_{ni} = \|\mathbf{q}_n[t] - \mathbf{q}_{\text{NFZ}}^i\| - R_{\text{NFZ}}$ denotes the distance from UAV $n$ to the boundary of NFZ $i$.  
The collision penalty is given by
\begin{equation}
P_{3} = \kappa_{3} \sum_{n=1}^{N} \sum_{n' = n+1}^{N}
\exp\left( \frac{d_{\min}-\|\mathbf{q}_n[t]-\mathbf{q}_{n'}[t]\|}{d_{\min}} \right).
\end{equation}
The NFZ-approaching reward is expressed as
\begin{equation}
 P_{4}=\sum_{n=1}^{N}\sum_{k=1}^{K}\sum_{i=1}^{I}\alpha_{n,k}[t]\cdot\mathbb{I}_{k,i}[t]\cdot \exp\big(\kappa_{4}(d_\text{NFZ}-D_{ni})\big),
    \label{eq:P4}
\end{equation}
where $\mathbb{I}_{k,i}[t]$ is an indicator function that equals 1 if MU $k$ is located inside NFZ $i$ at time slot $t$ (i.e., $||u_{k}[t]-q_\text{NFZ}^{i}||^{2}<(R_\text{NFZ}^{i})^{2}$), and 0 otherwise. It should be noted that the parameters $\kappa_1$, $\kappa_2$, $\kappa_3$, $\kappa_4$, and $w$ must be carefully tuned to ensure convergence and achieve the desired performance.
	
	\subsection{SAC-Based Algorithm}
\begin{figure}[t]
		\centering
		\includegraphics[width=\columnwidth]{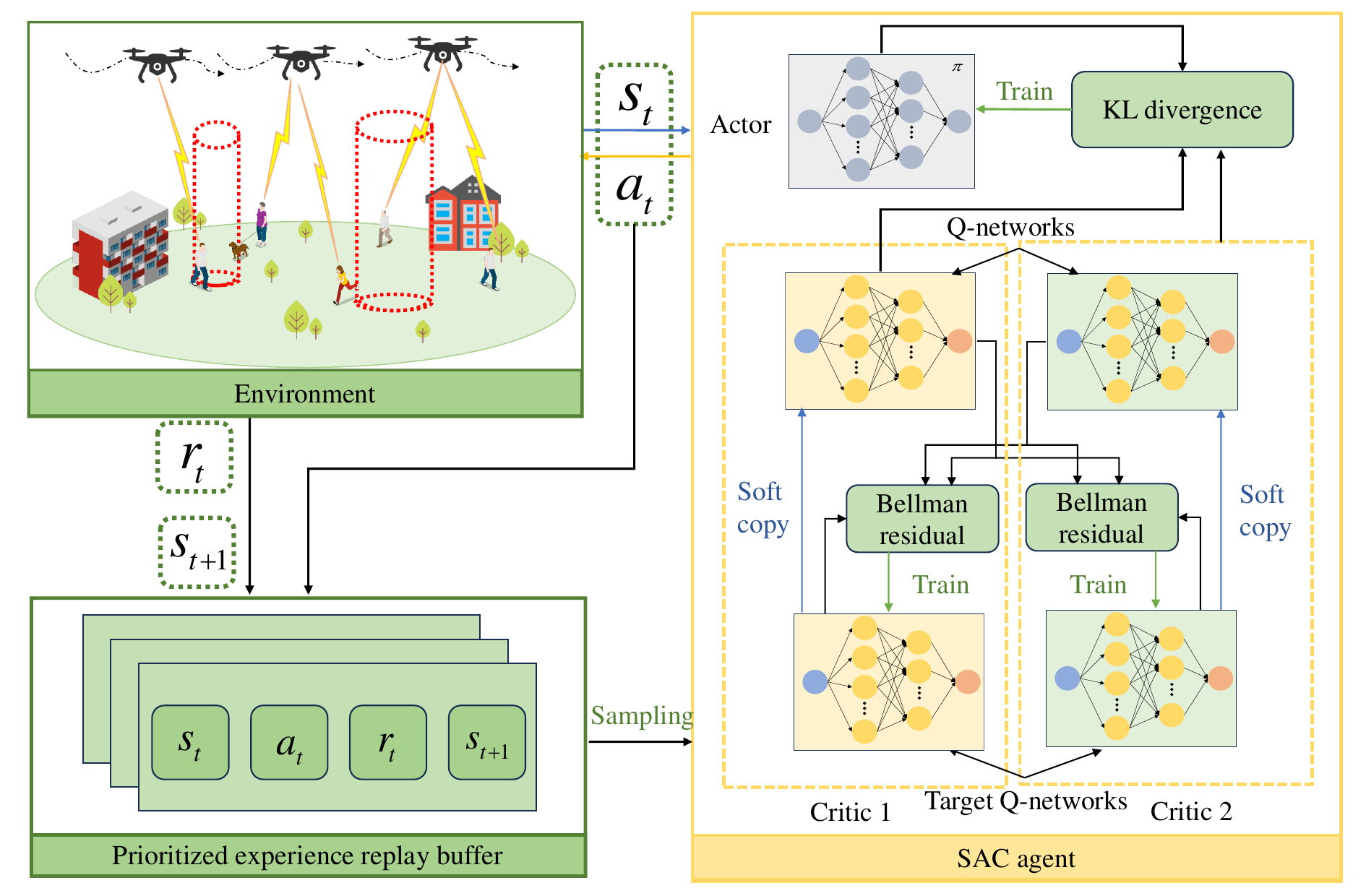}
		\caption{The SAC-PER framework.}
		\label{fig:sac-per}
	\end{figure}

	The architecture of SAC consists primarily of the following components: an actor network (policy network), critic networks (value networks), a temperature parameter \( \alpha \) that balances exploration and exploitation, and a replay buffer for experience storage and reuse. The training framework of SAC-PER is illustrated in Fig.~\ref{fig:sac-per}.
	
	\textit{1) SAC-based Framework:} Our UAV control problem involves (i) continuous action spaces, (ii) exploration under safety constraints, and (iii) the spatial mobility and requirement heterogeneity of MUs. SAC is well suited to such settings because it couples off-policy data efficiency with maximum-entropy exploration, yielding robust optimization in continuous domains. The entropy term discourages premature exploitation and improves robustness to reward shaping and variations in dynamics~\cite{Haarnoja2018SoftAO}. Concretely, SAC optimizes the entropy-regularized return

	\begin{equation}
    F(\pi) = \sum_{t=1}^{T} \mathbb{E}_{(s_t, a_t) \sim \rho_\pi} \left[\gamma^{t-1} r_t \right] 
        + \alpha \sum_{t=1}^{T} \mathbb{E}_{s_t \sim \rho_\pi} \left[ \mathcal{H}\left(\pi(\cdot|s_t)\right) \right],
\end{equation}
	where $\mathcal{H}(\pi(\cdot|s_t))$ denotes the entropy of the policy, defined as
	\begin{equation}
		\mathcal{H}(\pi(\cdot|s_t)) = - \mathbb{E}_{a_t \sim \pi(\cdot|s_t)} \left[ \log \pi(a_t|s_t) \right],
	\end{equation}
	which reflects the degree of randomness in action selection. The temperature parameter $\alpha$ regulates the trade-off between exploration and exploitation by adaptively tuning the stochasticity of the policy. To mitigate the overestimation bias in Q-value estimation, the SAC algorithm incorporates Twin Q-learning for updating the state-action value function. Specifically, the two Q-networks, denoted as $Q_{\theta_1}$ and $Q_{\theta_2}$, are used to compute the target Q-value as follows:
	\begin{equation}
		y_t = r_t + \gamma \left[ - \alpha \log \pi(a_{t+1} | s_{t+1}) + \min_{i=1,2} Q_{\theta_i}(s_{t+1}, a_{t+1}) \right],
		\label{eq:target}
	\end{equation}
	where $\gamma$ is the discount factor that controls the weight of future rewards, while the target value utilizes the minimum of the two Q-values to mitigate overestimation risk. The term $\log \pi(a_{t+1}|s_{t+1})$ acts as an entropy regularizer to enhance exploration. The Q-network loss function is defined as
	\begin{equation}
		F_Q({\theta_i}) = \mathbb{E}_{(s_t,a_t,r_t,s_{t+1}) \sim \mathcal{D}} \left[ \left(y_t - Q_{\theta_i}(s_t, a_t) \right)^2 \right], i=1,2,
		\label{eq:critic}\end{equation}
	where $\mathcal{D}$ is the replay buffer, which stores historical data for training. SAC uses a Gaussian policy, where the policy parameter \( \pi(a_t|s_t) \) is optimized by maximizing the expected advantage function, which involves optimizing the expected return and minimizing the value function bias. This objective can be expressed as
	\begin{equation}
		F_\pi(\phi) = \mathbb{E}_{a_t \sim \pi_\phi} \mathbb{E}_{s_t \sim \mathcal{D}} \left[ \alpha \log \pi_\phi(a_t|s_t) - Q(s_t, a_t) \right],
		\label{eq:policy}
	\end{equation}
	where \( \log \pi_\phi(a_t|s_t) \) regulates the policy entropy, encouraging sufficient exploration, while \( Q(s_t, a_t) \) represents the estimated expected return under the current policy.
	
	\textit{2) Temperature Parameter Adjustment:} The temperature parameter in SAC plays a critical role, as it governs the balance between exploration and exploitation by modulating the trade-off between reward maximization and policy entropy. In the early stages of training, SAC encourages extensive exploration by adopting a relatively large value of $\alpha$. As training advances, this value is gradually reduced, leading to more deterministic behavior and improved policy stability. To achieve a dynamic balance throughout training, SAC introduces an automatic temperature adjustment mechanism, which optimizes $\alpha$ by minimizing the following objective:
	\begin{equation} F(\alpha) = \mathbb{E}_{a_t \sim \pi
			_{\theta}}[-\alpha \log \pi_\theta(a_t|s_t) - \mathcal{H}_{min}], 
	\end{equation}
	where $\mathcal{H}_{min}$ denotes the minimum entropy constraint at
	each time slot. SAC updates $\alpha$ via gradient descent as follows:
	\begin{equation}
		\alpha \leftarrow \alpha - \lambda_{\alpha} \nabla_{\alpha} F(\alpha),
		\label{eq:alpha_adjust}
	\end{equation}
	where $\lambda_{\alpha}$ is the learning rate.
	
	\textit{3) Prioritized Experience Replay (PER):} In RL training, the traditional experience replay buffer uses a uniform sampling strategy. While this method improves data reuse, it overlooks the varying importance of individual experiences during policy updates, which can lead to suboptimal learning efficiency. To address this limitation, this paper employs PER to improve the training process of SAC. PER assigns a priority value to each sampled experience, thereby increasing the sampling probability of experiences considered more important. The temporal difference (TD) error quantifies the discrepancy between the predicted and target Q-values, computed as
	\begin{equation} \delta_t = \left| Q(s_t, a_t) - y_t\right|, 
		\label{eq:TD_error}\end{equation}
	a larger TD-error indicates a greater potential impact on policy improvement, and accordingly, such samples are replayed more frequently to accelerate learning convergence. Therefore, the probability of prioritized sampling $P(i)$ is defined as
	\begin{equation} P(i) = \frac{p_i^\beta}{\sum_j p_j^\beta}, 
		\label{eq:sample_pi}\end{equation}
	where $p_i = |\delta_i| + \epsilon$ denotes the priority assigned to sample $i$, and $\epsilon$ is a small positive constant introduced to avoid the assignment of zero priority, ensuring that all experiences have a non-zero chance of being sampled. The parameter $\beta$ serves as a control variable that adjusts the level of prioritization, balancing the impact of high-priority samples during the learning process. In addition, to mitigate sampling bias introduced by non-uniform sampling, PER incorporates importance sampling (IS) to adjust the sample weights, thereby promoting stable and unbiased training. The IS weight corresponding to sample $i$ is defined as
	\begin{equation} w_i = \left( \frac{1}{M} \cdot \frac{1}{P(i)} \right)^{\lambda},
		\label{eq:sample}
	\end{equation}
	where $M$ denotes the size of the replay buffer, and $\lambda$ is a tunable parameter that governs the strength of the importance sampling correction. Typically, $\lambda$ is annealed from 0 to 1 throughout the training process to gradually compensate for the bias while maintaining stability.
	
	By integrating PER into the SAC framework, the convergence efficiency is significantly improved, resulting in enhanced policy stability and adaptability in dynamic environments. This integration is particularly beneficial in RL scenarios involving complex tasks, including trajectory planning and power allocation. Algorithm \ref{SAC-PER} summarizes the proposed SAC-PER algorithm.

	\begin{algorithm}[t]
		\caption{Our proposed SAC-PER algorithm}
		\label{SAC-PER}
		\begin{algorithmic}[1]
			\STATE{Initialize environment.}
			\STATE{Initialize critic networks $\theta_i\ (i=1,2)$, actor network $\phi$.}
			\STATE{Initialize hyperparameters: entropy level $\mathcal{H}_{\min}$, learning rate, temperature parameter $\alpha$, discount factor $\gamma$, replay buffer $\mathcal{D} = \emptyset$.}
			\FOR{each episode}
			\FOR{each environment step}
			\STATE{Observe initial state $s_t$, select action $a_t$ according to current policy: $a_t \sim \pi_{\phi}(\cdot|s_t)$, execute action $a_t$.}
			\STATE{Transition to next state $s_{t+1}$, calculate reward $r_t$, store tuple $\{s_t, a_t, r_t, s_{t+1}\}$ in $\mathcal{D}$.}
			\IF{samples in $\mathcal{D}$ exceed buffer size}
			\FOR{$b \in \mathcal{B}_{\text{batch}}$}
			\STATE{Sample $i$ with probability $P_i$.}
			\STATE{Compute importance sampling by \eqref{eq:sample}.} 
			\STATE{Compute TD-error $\delta_i$ by \eqref{eq:TD_error}.}
			\STATE{Update priority $P_i$ by \eqref{eq:sample_pi}.}
			\ENDFOR
			\ENDIF
			\ENDFOR
			\FOR{each gradient step}
			\STATE{Compute target value using \eqref{eq:target}.}
			\STATE{Update critic networks by minimizing loss \eqref{eq:critic}}: 
			
			$\theta_i \leftarrow \theta_i - \lambda \nabla_{\theta_i} F_Q (\theta_i),  i \in \{1, 2\}.$
			\STATE{Update actor network by minimizing policy loss \eqref{eq:policy}}:
			
			$\phi \leftarrow \phi - \lambda \nabla_{\phi} F_\pi (\phi).$
			\STATE{Update temperature parameter $\alpha$ to adjust entropy \eqref{eq:alpha_adjust}.}
			\ENDFOR
			\ENDFOR
			\RETURN{Optimal policy $\pi^*$}
		\end{algorithmic}
	\end{algorithm}
	\subsection{Complexity Analysis}
	In the SAC-PER algorithm, the training time of the deep neural networks, specifically the actor and critic networks, constitutes the primary source of computational complexity. During the training process, each iteration operates on a mini-batch of samples, denoted by $\mathcal{B}_{\text{batch}}$. Accordingly, the computational complexity of performing gradient descent updates for both the critic and actor networks is determined by the size of $\mathcal{B}_{\text{batch}}$ and the respective network architectures.
	\begin{equation} \mathcal{O} \left(\mathcal{B}_{\text{batch}}  \left(\sum_{u=0}^{U-1} C_u C_{u+1} + \sum_{v=0}^{V-1} \hat{C}_v \hat{C}_{v+1}\right) \right), \end{equation}
	where \( C_u \) represents the number of neurons in the \( u \)-th layer of the actor network, and \( \hat{C}_v \) represents the number of neurons in the \( v \)-th layer of the critic network. This expression reflects the total computational cost involved in gradient calculations during each training step. The variables \( U \) and \( V \) correspond to the number of fully connected layers in the actor and critic networks, respectively. Therefore, the computational complexity of a single epsilon-step is characterized by the batch size $\mathcal{B}_{\text{batch}}$ and the model complexity, and is given by
	\begin{equation}
		\mathcal{O} \left( \text{TElen}\left(\mathcal{B}_{\text{batch}}  \left(\sum_{u=0}^{U-1} C_u C_{u+1} + \sum_{v=0}^{V-1} \hat{C}_v \hat{C}_{v+1}\right) \right) \right),
	\end{equation}
	where $\text{TElen}$ represents the total number of training episodes.
	
	Based on the analysis above, it is clear that the computational complexity of the SAC-PER algorithm mainly arises from the gradient computation and parameter updates during the training of the actor and critic networks. Specifically, the complexity associated with gradient backpropagation is directly related to the number of neurons in each layer of the deep neural networks. In practical applications, optimizing the network architecture, such as using layer reduction and parameter sharing techniques, can significantly alleviate the overall communication load.
	
	\section{SIMULATION RESULTS}\label{pro:s}
		In this section, the performance of the proposed SAC-PER algorithm is evaluated using the sum of priority rates as the main performance metric. To ensure a fair and comprehensive comparison, the following benchmark algorithms are considered:
	\begin{itemize}
		\item \textbf{PPO (Proximal Policy Optimization):} PPO has emerged as one of the most widely adopted and reliable DRL algorithms due to its robustness and rapid convergence properties~\cite{ppo1, ppo2}. In contrast to conventional methods, PPO employs a stochastic policy and leverages distribution-based policy updates, which contribute to improved training stability and sample efficiency. By effectively balancing the trade-off between exploration and exploitation, PPO substantially enhances the practical performance of DRL. However, PPO is sensitive to hyperparameter choices, and improper tuning can result in suboptimal performance.

		\item \textbf{DDPG:} DDPG is a DRL algorithm that integrates deterministic policy gradients and is specifically tailored for tasks characterized by high-dimensional state spaces and continuous action domains~\cite{ddpg}. Unlike stochastic methods, DDPG generates deterministic actions directly, thereby improving computational efficiency and decision-making precision. However, DDPG, while suitable for continuous control, tends to be unstable.

		\item \textbf{Fixed UAVs Scheme:} In this scheme, the UAVs are deployed at predetermined optimal locations and remain stationary throughout the mission, without undergoing any positional adjustments.
        
		\item \textbf{No-priority Scheme:} This scheme assumes  that all MUs belong to the same type, without considering differences in MU demands.
        
	\end{itemize}
    
	\subsection{Simulation Setup}

The simulation setup is as follows. The initial positions of both UAVs and MUs are randomly distributed. The UAVs operate at a fixed altitude of $100\ \mathrm{m}$ within a horizontal area of $[500, 500]\ \mathrm{m}$. The hyperparameters of the SAC algorithm are as follows~\cite{art9}: the episode length is set to 1000 steps, and the maximum number of training episodes is \( \mathrm{TE} = 500 \). The replay buffer stores up to $5 \times 10^4$ transitions. To improve sampling efficiency, PER is implemented with a priority exponent of $\beta_{\text{PER}} = 0.6$, an initial importance sampling weight of $\lambda = 0.4$, and a small constant $\epsilon = 10^{-6}$ to prevent zero sampling probabilities. The actor and critic networks are trained with learning rates of $3 \times 10^{-4}$ and $2 \times 10^{-4}$, respectively. The discount factor is set to $\gamma = 0.99$, while the target smoothing coefficient and entropy coefficient are chosen as $\tau = 0.05$ and $\alpha = 0.2$ to stabilize learning and encourage exploration. Each update uses a batch of $256$ samples. Both networks have two hidden layers with sizes $[256, 128]$ and employ the ReLU activation function.

The MUs move within this area following the RSSGM. The implementation is carried out in \text{Python 3.11.10}, with \text{PyTorch 2.3.1} as the core deep learning framework. To ensure the credibility of the results, we conducted three independent training runs under different random seeds. The simulation parameters are summarized in Table~\ref{tab:parameters}.

    \begin{table}[htbp]
    \renewcommand{\arraystretch}{1.2}
    \centering
    \caption{System Parameters}
    \begin{tabular}{@{}p{0.32\textwidth}@{\hspace{0.05\textwidth}}p{0.11\textwidth}@{}}
        \toprule
        Parameter & Value \\
        \midrule
        Time period $T$ & $200\ \mathrm{s}$ \\
        {Bandwidth $B_0$} & {$10^7\ \mathrm{Hz}$} \\
        Channel power gain $\rho$ & $-30\ \mathrm{dB}$ \\
        {Path-loss factor $\beta$} & {$10^{-2}$} \\
        Rician factor $\varsigma$ & $10$ \\
        Time slot $\delta$ & $1\ \mathrm{s}$ \\
        Noise power $\sigma^2$ & $-65\ \mathrm{dBm}$ \\
        UAV's flight altitude $H$ & $100\ \mathrm{m}$ \\
        NFZ's altitude $H_{\text{NFZ}}$ & $150\ \mathrm{m}$ \\
        Maximum UAV's acceleration $a_{\max}$ & $5\ \mathrm{m/s^2}$ \\
        Maximum UAV's speed $v_{\max}$ & $20\ \mathrm{m/s}$ \\
        Maximum UAV's transmission power $P_n^{\max}$ & $0.5\ \mathrm{W}$ \\
        {UAV battery capacity $E_n^{\text{UAV}}$} & {$6 \times 10^5\ \mathrm{J}$} \\ 
        Blade power $P_1$ & $59.03\ \mathrm{W}$ \\
        Hovering power $P_2$ & $79.07\ \mathrm{W}$ \\
        Rotor average speed $v_0$ & $3.6\ \mathrm{m/s}$ \\
        Blade tip speed $U_{\text{tip}}$ & $120\ \mathrm{m/s}$ \\
        Rotor area $A_0$ & $0.5030\ \mathrm{m^2}$ \\
        UAV's safety distance $d_{\min}$ & $3\ \mathrm{m}$ \\
        NFZ's boundary safety distance $d_{\text{NFZ}}$ & $10\ \mathrm{m}$ \\
        {NFZ radius $R_{\text{NFZ}}$} & {$ 30 \ \mathrm{m}$} \\ 
        {Priority weights $\zeta_z$} & {$[0.2, 0.3, 0.5]$} \\
        {Reward coefficients $\kappa_1, \kappa_2, \kappa_3, \kappa_4$} & {$10, 30, 40, -0.3$} \\
        Inertia coefficient $\eta$ & $0.85$ \\
        {MU's speed range $[v_u^{\min}, v_u^{\max}]$} & {$[2, 10]\ \mathrm{m/s}$} \\
        \bottomrule
    \end{tabular}
    \label{tab:parameters}
\end{table}

	\subsection{Performance Evaluation}
	\begin{figure}[t]
		\centering
		\subfigure[The convergence performance of different algorithms.]{
			\includegraphics[width=0.45\textwidth]{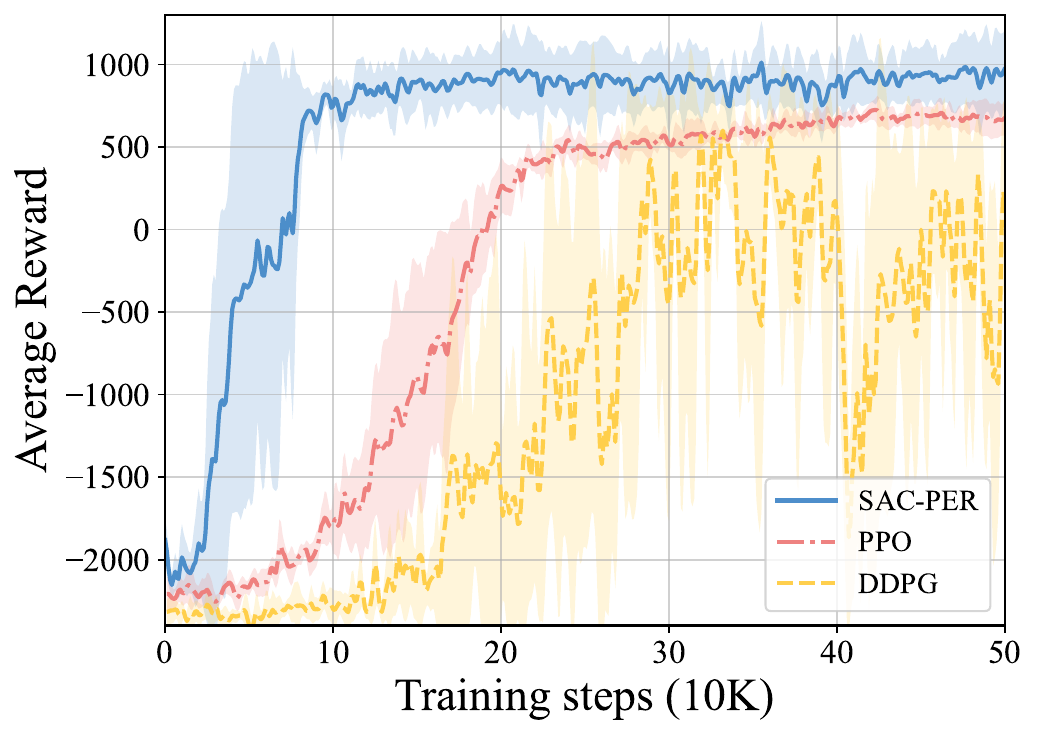}
		}
        \subfigure[The performance of SAC and SAC-PER algorithms.]{
			\includegraphics[width=0.45\textwidth]{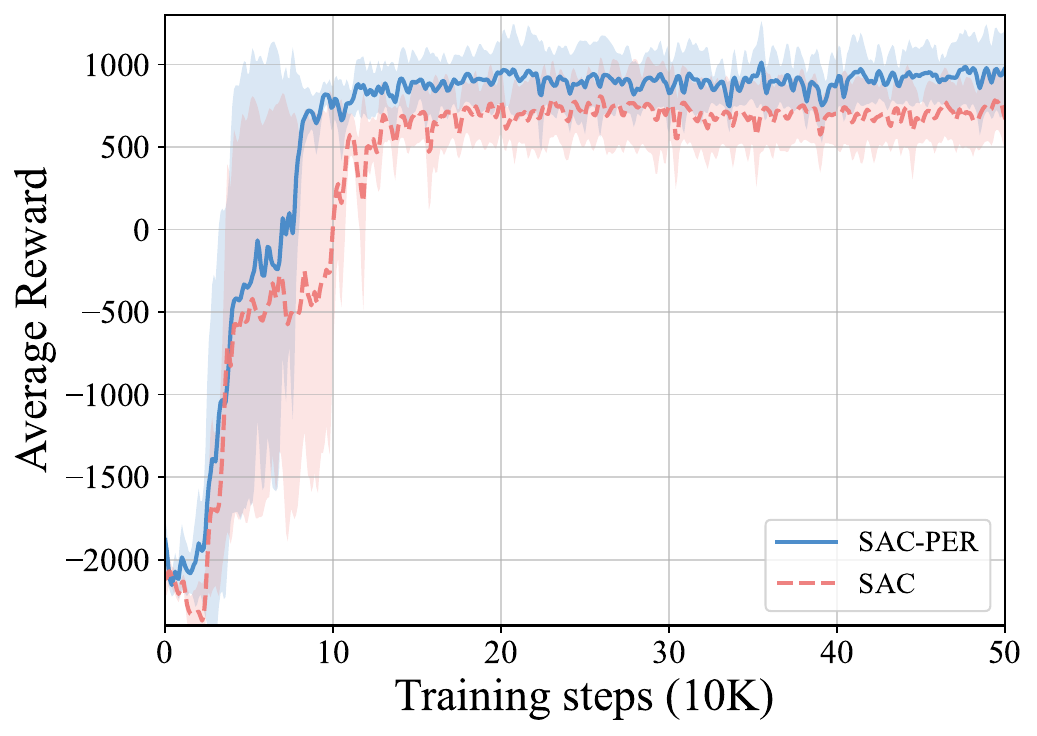}
		}
		\caption{The performance evaluation of the proposed algorithm.}
		\label{fig:convergen}
	\end{figure}

	\textit{1) Convergence:} Fig.~\ref{fig:convergen}(a) illustrates the convergence performance of SAC-PER through a comparative analysis with PPO and DDPG. It can be observed that all algorithms exhibit increasing reward values as training progresses, confirming the applicability of DRL techniques to complex communication environments. However, the convergence speed, stability, and reward performance of different algorithms vary significantly. As the agent gradually learns effective policies, SAC-PER starts to converge around 100K steps and subsequently stabilizes within a bounded range. In contrast, PPO exhibits a relatively slower convergence speed, and the final reward achieved is smaller than that of SAC-PER. Moreover, DDPG fails to exhibit effective convergence throughout the entire training period. In summary, SAC-PER demonstrates strong convergence efficiency and improved performance in complex DRL tasks, confirming its effectiveness and superiority over conventional RL methods.

    Fig.~\ref{fig:convergen}(b) illustrates the performance comparison between the SAC and the proposed SAC-PER algorithm. It can be observed that SAC-PER achieves significantly faster convergence than SAC and reaches a comparable performance level within fewer training steps. In addition, SAC-PER not only accelerates the convergence process but also exhibits higher stability and a slightly better final reward. These observations confirm that the prioritized sampling mechanism in PER enables the agent to focus more effectively on informative transitions, thereby enhancing both the learning efficiency and the overall performance. In the implementation, the mean performance across multiple runs is depicted by the solid curve, whereas the shaded region corresponds to the variance.

	\begin{figure}[t]
		\centering
		\subfigure[$N$=3, $K$=15, seed=2019.]{
			\includegraphics[width=0.22\textwidth]{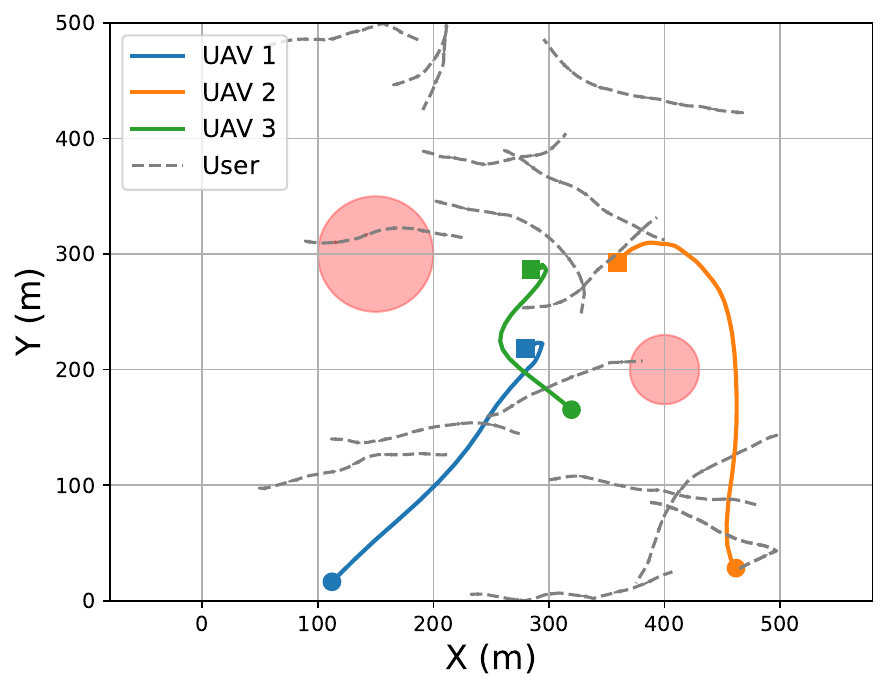}
		}
		\subfigure[$N$=3, $K$=15, seed=2020.]{
			\includegraphics[width=0.22\textwidth]{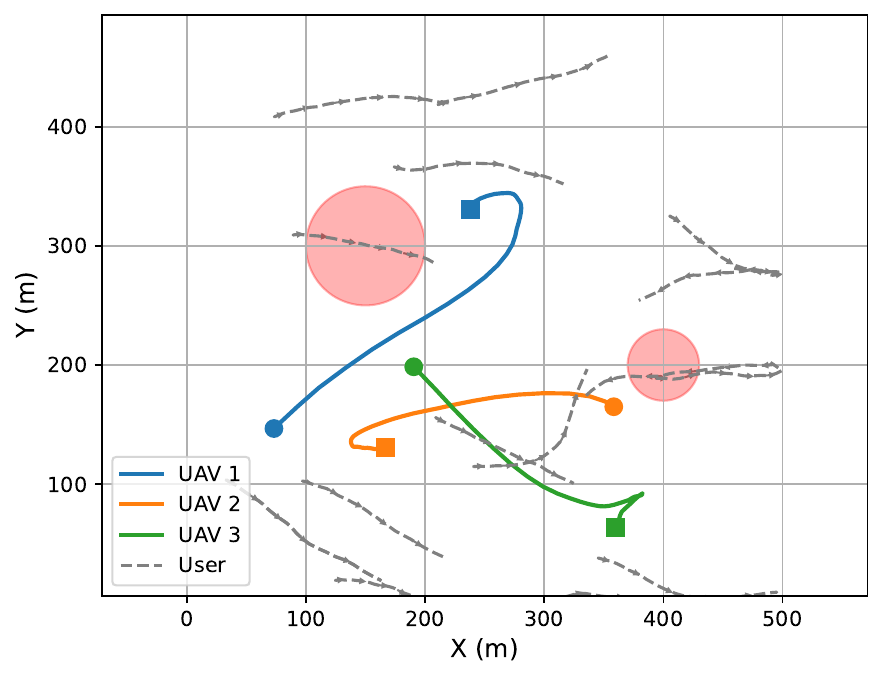}
		}
		\subfigure[$N$=5, $K$=25, seed=2019.]{
			\includegraphics[width=0.22\textwidth]{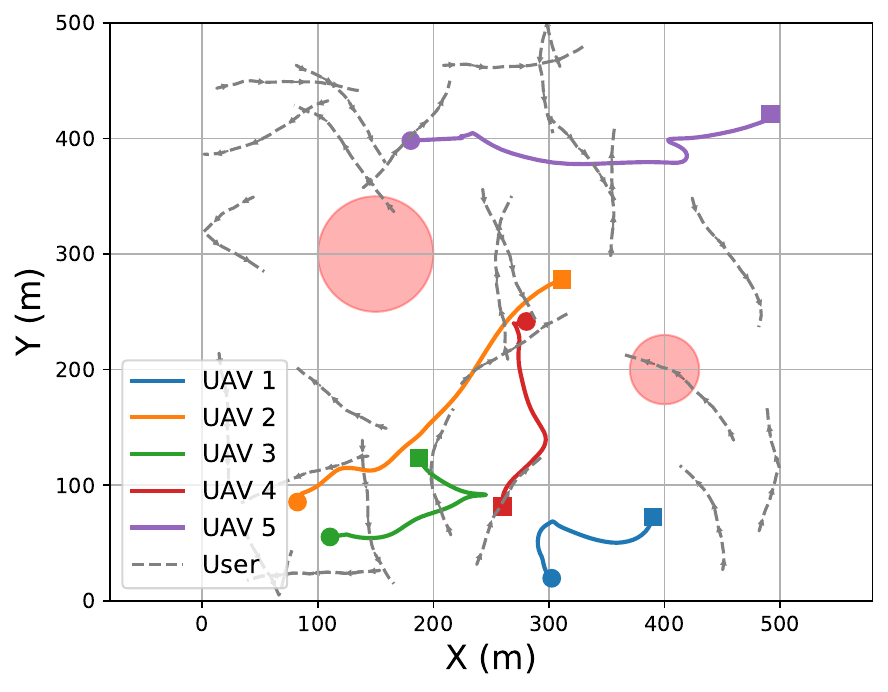}
		}
		\subfigure[$N$=5, $K$=25, seed=2020.]{
			\includegraphics[width=0.22\textwidth]{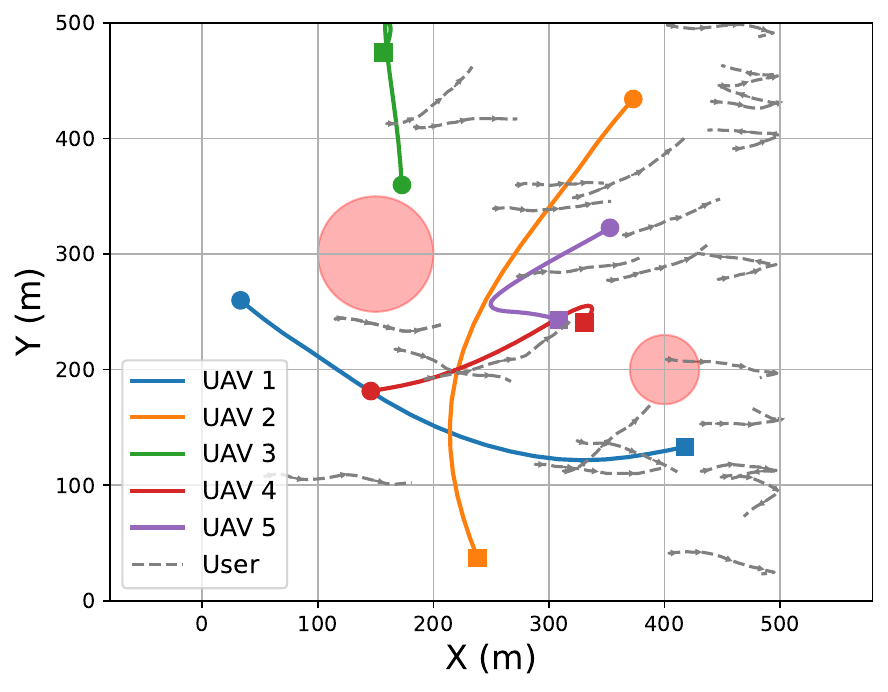}
		}
		\caption{Trajectories of UAVs and MUs in different scenarios.}
		\label{fig:trajectories}
	\end{figure}
	
	\textit{2) Trajectories:} Fig.~\ref{fig:trajectories} illustrates the trajectories of MUs and UAVs under two settings. As shown in Figs.~\ref{fig:trajectories}(a) and~\ref{fig:trajectories}(b), with 15 MUs and 3 UAVs, the UAVs progressively reposition toward clusters of active MUs, adapting to the MUs' spatial distribution. When certain MUs enter NFZs, the UAVs maneuver along the NFZ boundaries to preserve communication quality while remaining outside the restricted airspace. Figs.~\ref{fig:trajectories}(c) and~\ref{fig:trajectories}(d) depict the extended scenario introduced earlier. It can be observed that, while strictly avoiding NFZs, the UAVs gravitate toward MU-dense regions, thereby sustaining the aggregate communication rate.

	\begin{figure}[t]
		\centering
		\includegraphics[width=\columnwidth]{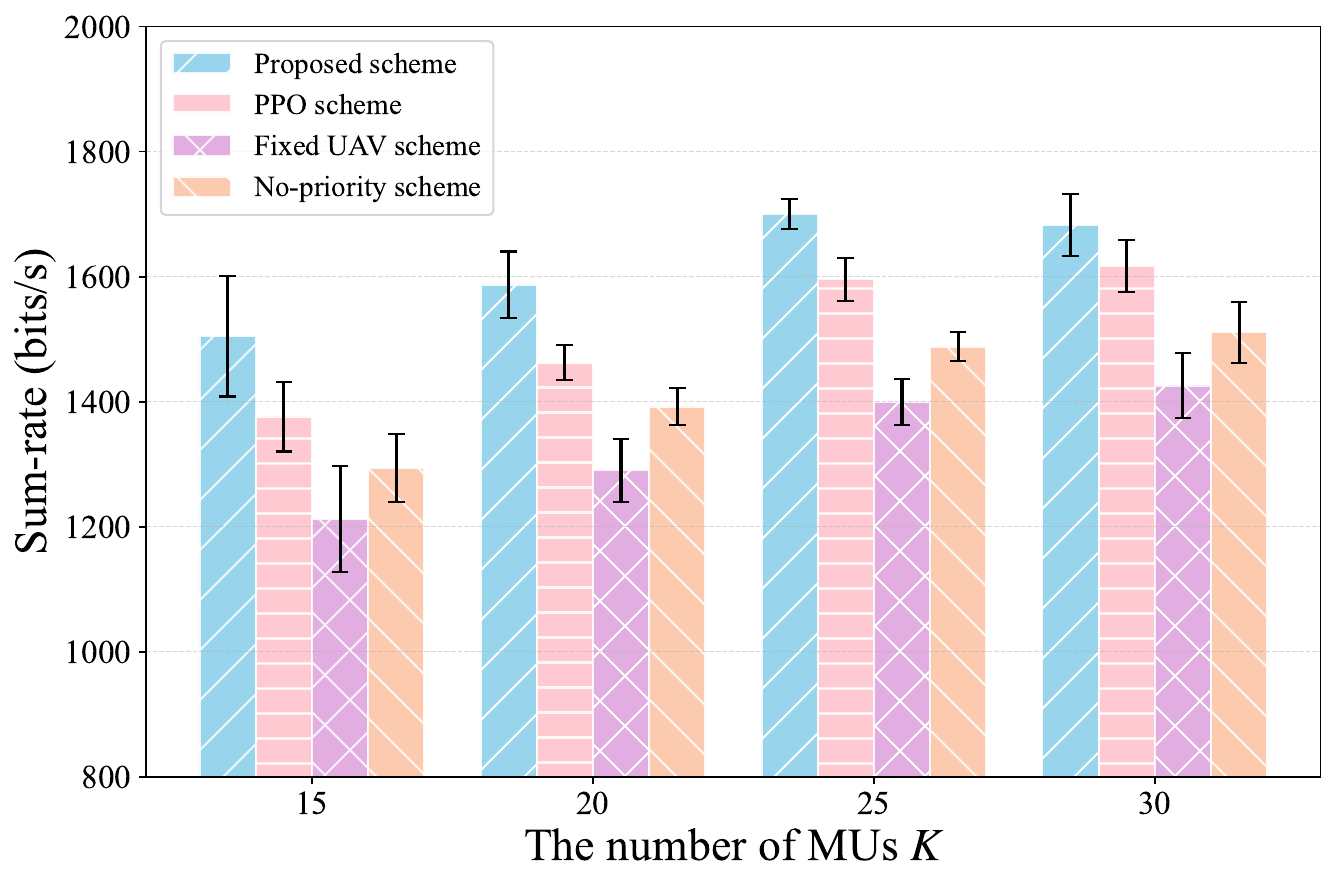}
		\caption{The sum-rate versus the number of MUs $K$.}
		\label{fig:user_num}
	\end{figure}
	\begin{figure}[t]
		\centering
		\includegraphics[width=\columnwidth]{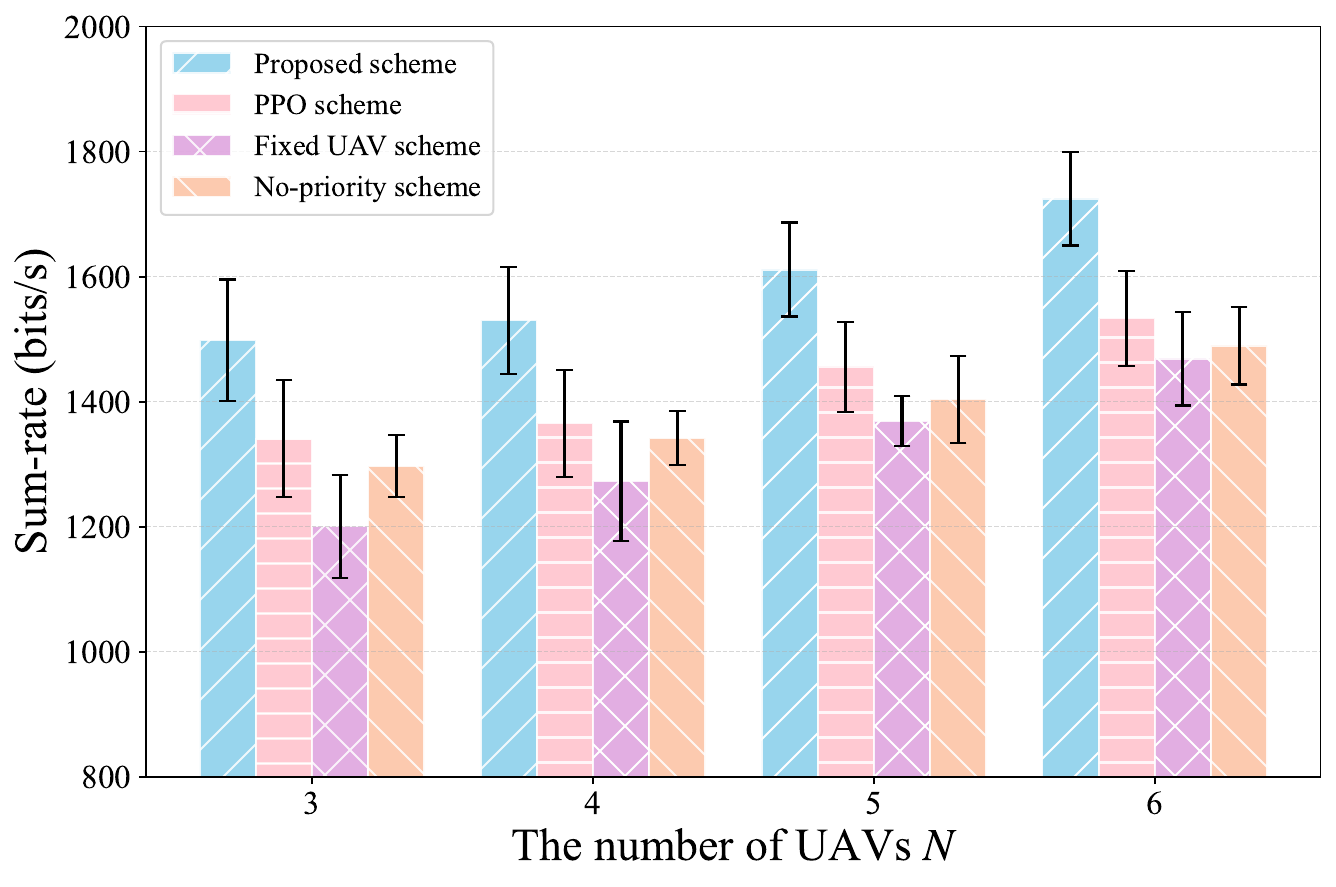}
		\caption{The sum-rate versus the number of UAVs $N$.}
		\label{fig:uav_num}
	\end{figure}
	
	\textit{3) Scalability: }To examine the scalability of the proposed scheme, we evaluate its performance under different MU numbers and compare it with baseline methods. As shown in Fig.~\ref{fig:user_num}, when the number of MUs is small, all schemes achieve higher sum-rates because each MU can be allocated a larger share of the total transmit power, which enhances link quality. As the MU population grows, however, the transmit power must be divided among more MUs, leading to reduced per-MU SINR. As a result, the increase in the overall sum-rate slows down and eventually tends to saturate. 
    
   As depicted in Fig.~\ref{fig:uav_num}, the achievable sum-rate increases with the number of UAVs. This is because denser UAV deployment reduces the average UAV-MU distance and enhances spatial diversity, thereby improving link quality and spectrum utilization. Consequently, the system exhibits a notable performance gain. Furthermore, the proposed scheme consistently outperforms the benchmark approaches in all scenarios, highlighting its robustness and scalability in dynamic multi-MU environments.

\begin{figure}[t]
		\centering
		\includegraphics[width=\columnwidth]{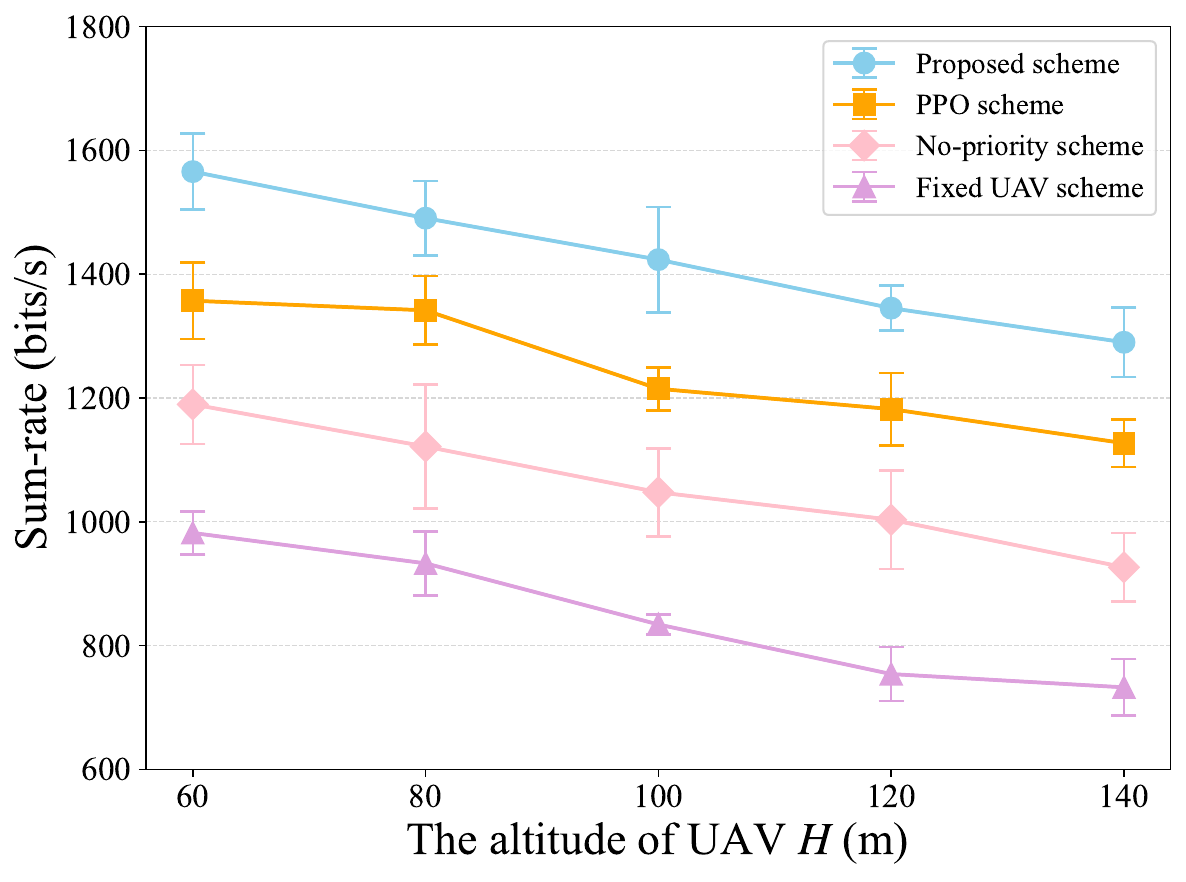}
		\caption{The sum-rate versus the altitude of UAV.}
		\label{fig:uav_height}
	\end{figure}
	\textit{4) Sensitivity: }
   To compare the performance of the proposed scheme with conventional methods, we further evaluate the total transmission rates achieved under different flight altitudes of the UAV, as illustrated in Fig.~\ref{fig:uav_height}. When the UAV's altitude is set to 60 $\mathrm{m}$, the proposed scheme exhibits a clear performance advantage. This improvement is primarily due to the reduced distance between UAVs and MUs, which enhances channel quality and thereby increases the sum-rate. In contrast, fixed-position UAVs lack adaptability to dynamic environmental conditions, leading to suboptimal performance compared to the proposed adaptive approach. Moreover, since conventional schemes do not consider communication priorities, resource allocation may be inefficient, with UAVs potentially serving lower priority MUs, which further degrades system performance. In contrast, the proposed SAC-PER method optimizes in continuous spaces, offering a more flexible and efficient solution, demonstrating its superiority under time-varying channel conditions.
   
	\begin{figure}[t]
		\centering
		\subfigure[The sum-rate versus the number of NFZs.]{
			\includegraphics[width=0.45\textwidth]{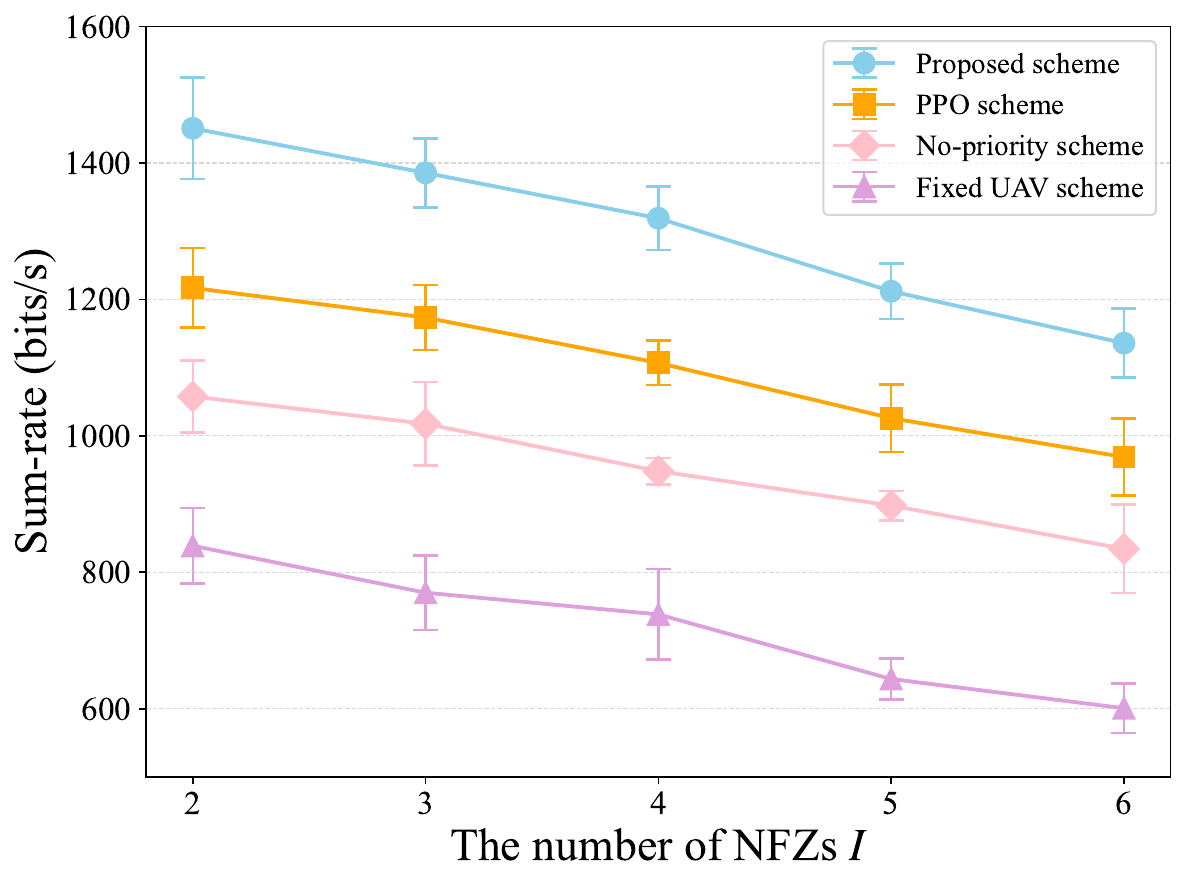}
		}
        \subfigure[The sum-rate versus the radius of NFZs.]{
			\includegraphics[width=0.45\textwidth]{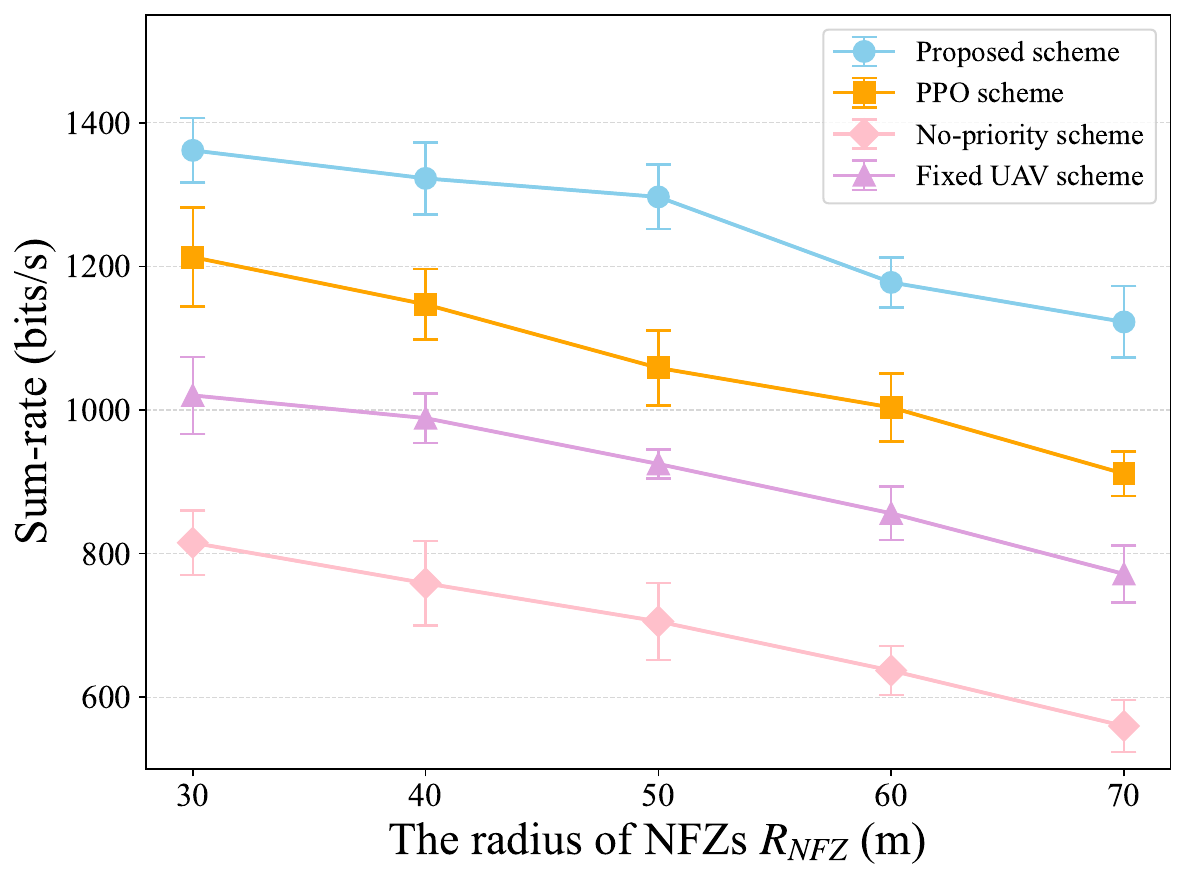}
		}
		\caption{The impact of NFZ spatial constraints on the system sum-rate.}
		\label{fig:nfz_impact}
	\end{figure}
    
	Fig.~\ref{fig:nfz_impact} evaluates the impact of spatial constraints on system performance by varying the number and radius of NFZs. A consistent degradation in the sum-rate is observed across all schemes in both scenarios. Specifically, as shown in Fig.~\ref{fig:nfz_impact}(a), a higher density of NFZs reduces the feasible flying space and increases the probability of NLoS links. Similarly, Fig.~\ref{fig:nfz_impact}(b) illustrates that expanding the NFZ radius further shrinks the available airspace, compelling UAVs to bypass spatially optimal communication spots to strictly satisfy collision-avoidance constraints. Fundamentally, whether through increased density or enlarged boundaries, the expanded restricted regions force UAV spatial displacements that directly compromise the overall transmission performance.

    \begin{figure}[t]
		\centering
		\includegraphics[width=\columnwidth]{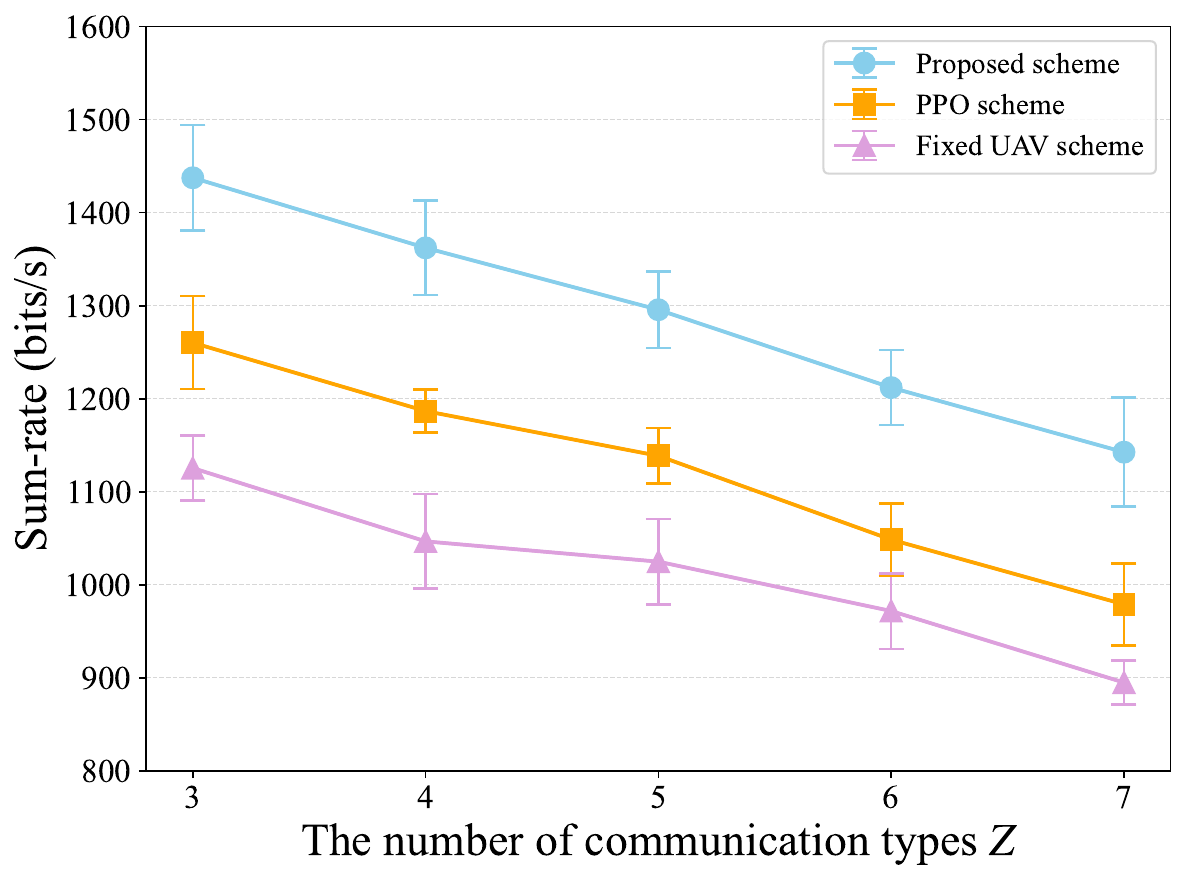}
		\caption{The sum-rate versus the number of communication types $Z$.}
		\label{fig:type_comm}
	\end{figure}

As depicted in Fig.~\ref{fig:type_comm}, the proposed scheme achieves the highest total transmission rate compared to the baselines across varying numbers of communication types. However, an overall decline in the unweighted sum-rate is observed as the number of communication types grows. An increased variance in priority classifications inherently drives the DRL agent to favor high-priority MUs. By skewing spatial proximity and transmit power toward these critical users to maximize the weighted reward, the UAVs deviate from the spatial optimal locations required to maximize the overall average channel gain, thereby reducing the unweighted sum-rate.

	\section{Conclusion}
	In this paper, we have investigated a multi-UAV-assisted communication system. Unlike prior studies, the proposed framework has allowed MUs to enter NFZs and has assigned different communication priorities according to their heterogeneous service requirements. To address the challenges posed by user mobility, diverse communication demands, and NFZ constraints, we have jointly optimized the transmit power allocation, the trajectory planning of UAVs, and the binary association factor between MUs and UAVs, aiming to maximize the weighted sum-rate based on MU priority levels. Due to the strong coupling among these variables, we have employed SAC-PER and an iterative algorithm to solve the original optimization problem. Simulation results have verified that the proposed method has significantly enhanced the sum-rate and has achieved superior performance in terms of convergence speed and training stability compared with baseline algorithms. Future work may explore joint 3D path planning, decentralized multi-agent DRL for more scalable UAV cooperation, and the integration of strict service-specific QoS constraints to further enhance system reliability.

	\bibliographystyle{IEEEtran}
	\bibliography{reference.bib} 
	
\end{document}